\documentclass[12pt,preprint, appendixfloats]{aastex}
\usepackage{color}

\usepackage{changes}

\newcommand{\nv}{N\,{\footnotesize V}}
\newcommand{\civ}{C\,{\footnotesize IV}}
\newcommand{\siiv}{Si\,{\footnotesize IV}}
\newcommand{\mgii}{Mg\,{\footnotesize II}}
\newcommand{\mgiis}{Mg\,{\footnotesize II} $\lambda$2796}
\newcommand{\mgiie}{Mg\,{\footnotesize II} $\lambda$2803}

\newcommand{\Aliii}{Al\,{\footnotesize III}}

\newcommand{\oii}{[O\,{\footnotesize II}]}

\newcommand{\nii}{[N\,{\footnotesize II]}}

\newcommand{\hr}{H$\gamma$}
\newcommand{\hb}{H$\beta$}
\newcommand{\ha}{H$\alpha$}
\newcommand{\heitnff} {He\,{\footnotesize I} $\lambda$2945}
\newcommand{\hei}{He\,{\footnotesize I*}}
\newcommand{\heitoen} {He\,{\footnotesize I*} $\lambda$3189}
\newcommand{\heiteen} {He\,{\footnotesize I*} $\lambda$3889}
\newcommand{\heiozetz}{He\,{\footnotesize I*} $\lambda$10830}

\newcommand{\oiii}{[O\,{\footnotesize III}]}
\newcommand{\ovi}{O\,{\footnotesize VI}}
\newcommand{\oiiiftft}{[O\,{\footnotesize III}] $\lambda$4353}

\newcommand{\sii}{[Si\,{\footnotesize II}]}

\newcommand{\feii}{Fe\,{\footnotesize II}}
\newcommand{\fei}{Fe\,{\footnotesize I}}

\newcommand{\kmps}{$\rm km~s^{-1}$}
\newcommand{\cf}{$C_f$}
\newcommand{\tauv}{$\tau(v)$}

\def\gtsima{$\; \buildrel > \over \sim \;$}
\def\simgt{\lower.5ex\hbox{\gtsima}}

\begin{document}
\title{\bf Broad Emission and Absorption Line Outflows in the Quasar SDSS J163345.22+512748.4 }
\author{Bo Liu\altaffilmark{1,2}, Hongyan Zhou$^\dagger$\altaffilmark{1,2}, Xinwen Shu$^\ddagger$\altaffilmark{3}, Shaohua Zhang\altaffilmark{2},Tuo Ji\altaffilmark{2}, Xiang Pan\altaffilmark{2},Peng Jiang\altaffilmark{2}}
\altaffiltext{1}{Key Laboratory for Research in Galaxies and Cosmology, University of Science and Technology of China, 96 Jinzhai Road, Hefei, Anhui, 230026}
\altaffiltext{2}{Polar Research Institute of China, 451 Jinqiao Road, Shanghai, 200136, China; $^\dagger$Correspondence: zhouhongyan@pric.org.cn}
\altaffiltext{3}{Department of Physics, Anhui Normal University, Wuhu, Anhui, 241000, People Republic of China; $^\ddagger$Correspondence: xwshu@mail.ahnu.edu.cn 0000-0002-7020-4290}
\begin{abstract}

We present a detailed study of the optical and NIR emission and absorption line spectrum of the quasar SDSS J163345.22+512748.4. We discovered on the newly acquired NIR spectrum a highly meta-stable neutral helium broad
absorption line (BAL) \heiozetz\ with a width of $\sim$ 2000 \kmps\ and a
blueshift of $\sim$ 7000 \kmps\ in the velocity space. The BAL system is also significantly
detected in \mgii\ and \heiteen. We estimate a column
density of $(5.0 \pm 1.7) \times 10^{14}$ cm$^{-2}$ for the HeI*(2~$^3$S) level, and infer an ionization
parameter of $U_{A} = 10^{-1.9\pm 0.2}$ for the BAL outflow assuming
that the BAL region is thick enough for a full development of an
ionization front. The total column density of the BAL outflow is
constrained in the range  N$\rm _{H}$ $\sim$ 10$^{21}$-10$^{21.4}$ cm$^{-2}$. We also found that the bulk of both MgII and UV
FeII, as well as H$\alpha$
broad emission lines (BELs) are blueshifted with a velocity of $\sim$
2200 \kmps\ with respect to the quasar systemic redshift.
We constrain that the blueshifted BEL region has a
covering factor $C_{f}\approx 16\%$, a density n$\rm _{H}$  $\sim
$  10$^{10.6}$-10$^{11.3}$ cm$^{-3}$, a column density N$\rm _{H}\gtrsim
10^{23}$ cm$^{-2}$, and an ionization parameter $U_{E}\sim
10^{-2.1}-10^{-1.5}$.
The outflow gas is located at $\sim$0.1 pc away from the central ionization source,
at a scale comparable to the BLR.
A toy kinetic model has been proposed to reproduce the profile of MgII
BEL well if assuming a partial obscured axisymmetric geometry of the outflow with a radial
velocity as observed by the BALs.

\end{abstract}
\section{Introduction}
Outflows in active galactic nuclei (AGNs) play an important role in galaxy evolution. Recent studies indicate that the outflow is regulated by the accretion process (Sulentic et al. 2000; Leighly \& Moore 2004; Richards et al. 2011; Wang et al. 2011; Marziani \& Sulentic 2012).
By carrying away angular momentum,
the outflowing gas is crucial to maintain the accretion onto the central black hole (BHs) (Sulentic et al. 2000; Higginbottom et al. 2013; Feruglio et al. 2015; Fontanot et al. 2015), regulating the growth of the central supermassive BHs. Moreover, outflows are considered to be able to affect star formation in the host galaxies (Silk \& Rees 1998).
As one of important phenomena in the quasars, outflows leave prominent imprints in the quasar spectra,
such as blueshifted broad absorption lines (BALs; Weymann et al. 1991), as well as broad emission lines (BELs; Gaskell 1982).
To date, our study and understanding of the outflows are mainly based on the analysis of BALs and/or BELs.

BALs appear in the spectra of 10-15\% optically selected quasars. These quasars often show absorptions from both high and low ionization ions, such as \nv, \civ, \siiv, \ovi, \Aliii\ and \mgii (Hall et al. 2002; Tolea et al. 2002; Hewett \& Foltz 2003; Reichard et al. 2003; Trump et al. 2006; Gibson et al. 2009; Zhang et al. 2010, 2014). Studies of BALs can  place constraints on the physical properties of the outflows, which are helpful to understand the connection between the evolution of SMBHs and their host galaxies.
However, due to the single line of sight, the covering factor which is an important parameter of the BAL outflows, is difficult to be determined for an individual quasar. For most of BAL quasars,
the covering factor of outflows is usually derived in a statistical way from a sample of sources,
resulting in that the estimation for other properties may be not reliable.

As another important feature of outflows, the blueshifted BELs was first detected in the
high-ionization lines (e.g., \civ, Gaskell 1982; Wilkes 1984). The blueshifted BELs are
difficult to reconcile with gravitationally-bound BELR models, but can be considered as a
signature of outflowing gas (Gaskell 1982; Marziani et al. 1996; Leighly 2004; Wang et al. 2011).
Recently, blueshifted BELs have also been found in the low-ionization lines, such as \mgii, which
can be interpreted as the signature of a radiation-driven wind or outflow (Marziani et al. 2011).
Different from the BALs, the integral flux of blueshifted BELs can reflect the global properties of
outflowing gas. The equivalent widths (EWs) and line ratios can be used imposing strong constraints on the
density, ionization state, and geometry of the line emitting gas (Liu et al. 2016). However, in most
of the quasars with blueshifted BELs, the blueshifted BELs are always blended with the normal BELs
emitted from the broad line region (BLR) and the decomposition between them is a challenging task.

This paper presents a detailed emission line and absorption line analysis of SDSS J163345.22+512748.4 (hereafter SDSS J1633+5127), a type-1 quasar at z = 0.6289 with outflows revealed in both blueshifted BELs and BALs. Since its \mgii\ emission line is dominated by the blueshifted BELs, the uncertainty of the decomposing them from normal BELs is small. Besides \mgii, UV \feii\ and \ha\ also show similar blueshifted BEL components. These blueshifted lines can be considered emitted from outflows, for which
the properties can be inferred from the EWs and line ratios of BELs.
Combined with the properties of BALs, we provide new insights into outflowing gas.
The paper is organized as follows. The observational data are described in Section 2,
and further analysis is shown in Section 3. In Section 4, we give our discussions on the results.
In this paper, the cosmological parameters $H_0=70$~km~s$^{-1}$~Mpc$^{-1}$, $\Omega_{\rm M}=0.3$, and
$\Omega_{\Lambda}=0.7$ have been adopted throughout this paper.

\section{Observation and Data Reduction}

\begin{figure}[ht]
\epsscale{1.1}
\plotone{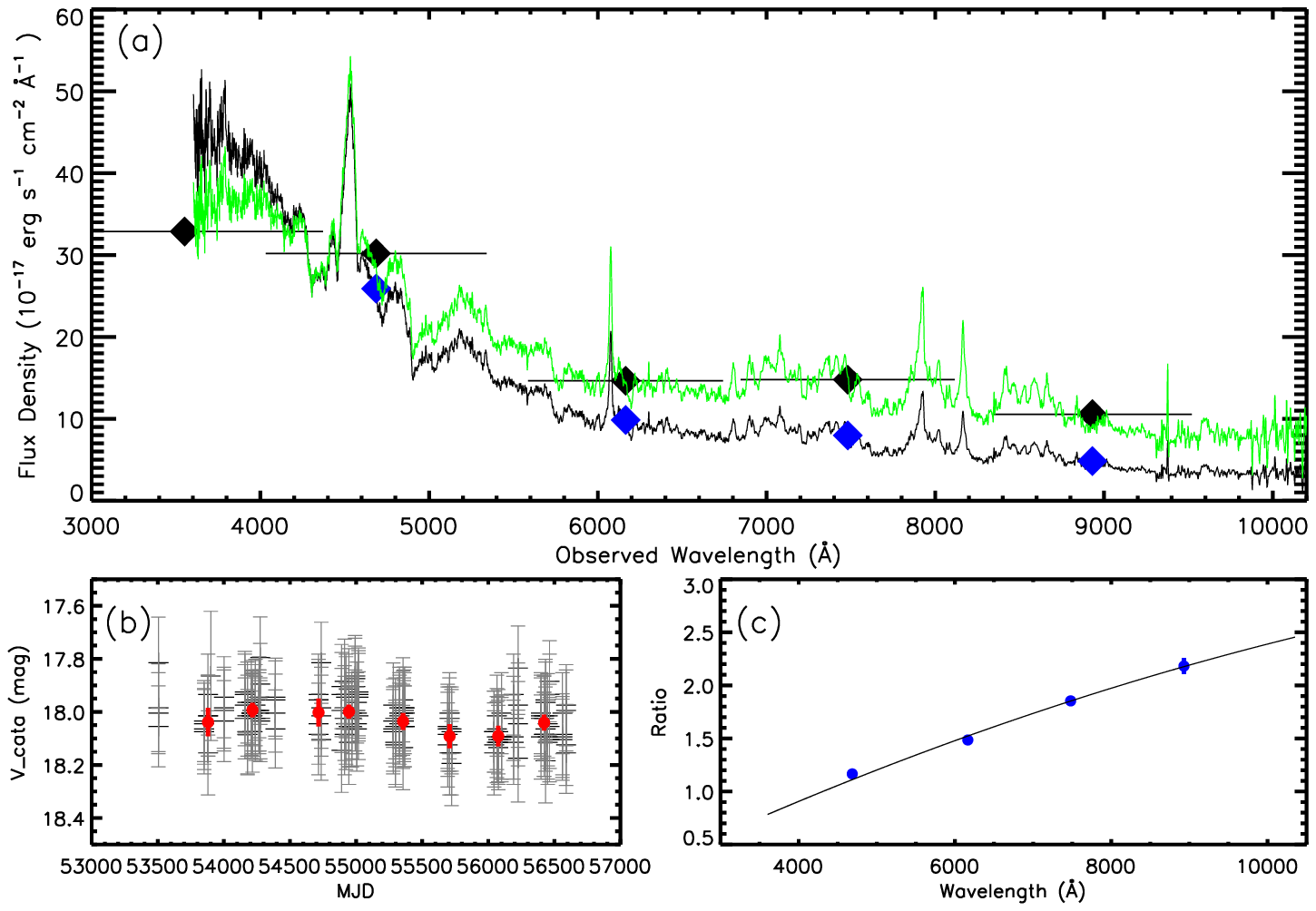}
\caption{\textbf{Panel (\textit{a})}: The observed spectra and photometry of SDSS  J1633+5127.
The BOSS spectrum (MJD 56191) are displayed by  black curve. For comparison, we plot the
SDSS five bands photometry (MJD 51948) and the BOSS spectral synthetic magnitude at \textit{g, r, i, }
and \textit{z} bands by black and blue diamonds, respectively.
The recalibrated BOSS spectrum is present in green. \textbf{Panel (\textit{b})}:  The light curve of SDSS  J1633+5127 at V band monitored by the
Catalina Sky Survey. The red dots represent the mean magnitude for each season.
The intrinsic source variability is about 0.06 magnitude
in 6.5 years in the rest-frame, which indicates the difference between the BOSS spectrum and SDSS photometry is likely
due to the spectrophotometric calibration uncertainty. \textbf{Panel (\textit{c})}:
The correction curve between the SDSS photometry and BOSS spectrum. The blue circles present the ratios of
SDSS photometry to BOSS spectrum in the \textit{g, r, i, z} bands.
The correction curve (grey) was fitted with 2 order polynomial.
}
\label{f1}
\end{figure}

   SDSS J1633+5127 was imaged by the SDSS on February 8, 2001. The point-spread function magnitudes measured from the images are $18.59 \pm 0.04$, $18.04 \pm 0.01$, $18.23 \pm 0.01$, $17.80 \pm 0.01$, $17.76 \pm 0.02$ at u, g, r, i, and z bands,
   respectively, which are shown with black diamonds in Fig.\ref{f1} (a).
   The optical spectrum of SDSS J1633+5137 was observed by the Baryon Oscillation Spectroscopic Survey (BOSS; Dawson et al. 2013) on October 23, 2011, for which spectrographs (Smee et al. 2013) can cover a wavelength range of 3600-10500 \AA.

The spectrum we used was extracted from the BOSS Date Release 10 (DR10; Ahn et al. 2014). After correcting
for the Galactic reddening of $E(B - V) = 0.051$ (Schlafly \& Finkbeiner 2011), the spectrum
is presented by black curve in the panel (a) of Fig.\ref{f1}.
The comparison with SDSS photometry clearly indicates that the spectrum has a bluer continuum slope and lower flux density
than the photometry at longer wavelengths.
We also calculate the spectral synthetic magnitudes at the g, r, i,and z bands,
which are shown in blue diamonds.
The later three magnitudes are even $\sim 1$ mag lower than the photometry.
This difference is possibly due to the BOSS spectrophotometric calibration uncertainty
or variability in the 6.5 rest-frame years between the two observations.

The Catalina Surveys Data Release 2\footnote{The Catalina Web site is http://nesssi.cacr.caltech.edu/DataRelease/. } (Drake et al. 2014) gives us an opportunity to clarify this issue. SDSS J1633+5137 is monitored for eight observing seasons, beginning in April 10, 2005. Each observing season, spanning from October to April next year, contains about 50 times photometric observations.
Since SDSS J1633+5137 is faint, the individual photometric error is large (about 0.5 mag) and
some observed magnitudes have large offsets from their neighbouring data likely due to noise fluctuations.
To display the light curve clearly, the photometric data in one observing season are combined and presented in the
panel (b) of Fig.\ref{f1}, which show very weak long-term variability with large measurement errors.
The intrinsic variability amplitude $\sigma_V=0.06~\rm mag$ ($=\sqrt{\Sigma_V^2-\xi^2}$; Ai et al. 2010)
is much smaller than the offset between the SDSS photometry and spectrum.
This suggests that the difference between the SDSS photometry and spectrum is more likely
to be caused by the spectrophotometric calibration uncertainties.
Thus, we attempted to use a 2-order polynomial to fit the flux ratios between the SDSS photometry and
spectral synthetic magnitudes
in g, r, i, and z bands,
which are shown in the panel (c) of Fig.\ref{f1}.
Using the fitted flux ratio at each wavelength bin,
we then scaled the spectrum to match the SDSS photometry
to obtain a recalibrated spectrum, which is shown with green in Fig.\ref{f1} (panel (a)).

At the infrared bands, we collected infrared photometric data of SDSS J1633+5137 from
the two micron all sky survey (2MASS; Skrutskie et al. 2006) and
the Wide-field Infrared Survey Explorer (WISE; Wright et al. 2010).
Meanwhile, we observed the near-infrared (NIR) spectra of SDSS J1633+5137
using the TripleSpec (Wilson et al. 2004) on the 200-inch Hale telescope at Palomar Observatory.
Four exposures of 300s each were taken in an A-B-B-A dithering mode with the primary configuration
of the instrument. A 1.1\arcsec\ slit was chosen to match the seeing. TripleSpec NIR spectrograph provides
simultaneous wavelength coverage from 0.9 to 2.46 microns at a resolution of 1.4 - 2.9 \AA.
The raw data were processed using IDL-based Spextool software (Cushing et al. 2004).
There are two gaps in the infrared spectrum around 1.35 microns and 1.85 microns due to the
effect of the atmosphere transmissivity.
Fortunately, the redshifted \ha\ emission line is detected with the TripleSpc at J-band.

\begin{figure}[ht]
\epsscale{1.1}
\plotone{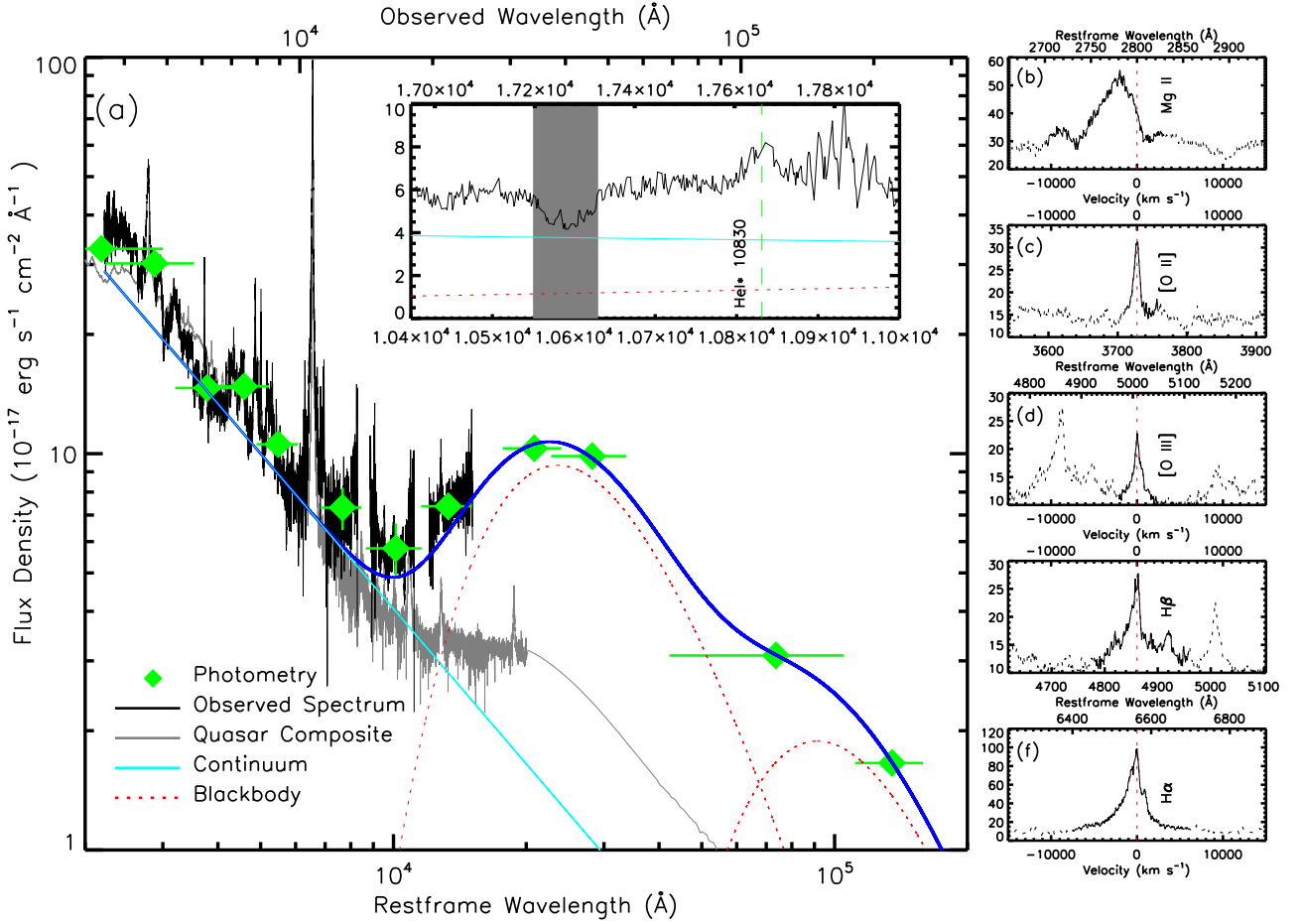}
\caption{\textbf{Panel (\textit{a})}: The UV to mid-infrared spectra and SED of SDSS J1633+5137 in the
quasar's rest-frame. The spectrum and photometry are present with black curve and green diamonds, respectively.
We modelled the broadband SED with a power law (cyan solid line) and two black bodies (red dotted line).
The sum of all modelled components (blue solid line) can roughly reproduce the continuum.
Compared to the quasar composite spectrum (gray), the excess in NIR bands implies that this source is
a BAL quasar which is confirmed in the NIR spectrum.
  \textbf{Panel (\textit{b})}: The spectrum near the \mgii\ emission line.
  Compare to the intrinsic wavelength (dotted line), \mgii\ of SDSS J1633+5137 is obviously blueshifted.
  \textbf{Panel (\textit{c,d,e})}: The spectrum near the \oii, \oiii, \hb, and \ha\ emission lines.
  All these lines converted from the redshift corresponds to the intrinsic wavelength, which indicates
  that the redshift is reliable.
}
\label{f2}
\end{figure}

After masking the bad and skyline seriously polluted pixels, we created a new spectrum by combining
the recalibrated optical spectrum with the TripleSpec NIR spectrum for the following analysis.
The systemic redshift of $z = 0.6289\pm0051$ reported from Paris et al. (2014) is consistent with that
derived from the narrow \oii, \oiii\ lines, and the peak of broad \hb, \ha\ lines. However, different from these lines, the \mgii\ shows blueshifted profile with the blueshifted velocity for the peak emission of about 2000 \kmps.  After being converted to the quasar rest-frame, the spectrum and the spectral energy distribution (SED) from ultraviolet (UV) to mid-infrared (MIR) from the SDSS, 2MASS, and WISE are shown in black curve and green points in the panel (a) of Fig.\ref{f2}.
The broad band SED of SDSS J1633+5137 is decomposed into a power law with
index of -1.3 (cyan) and two black bodies with a temperature of 1232 K and 312 K (red dotted), respectively.
Compared to the quasar composite spectrum (Zhong et al. 2010),  the SED of SDSS J1633+5137 shows
clear excess in the NIR bands.
As a common feature of BAL quasars where strong hot dust emission was found (Zhang et al. 2014),
this excess may hint at the existence of BALs in the spectrum.
Indeed, as shown in the inset panel of Fig.\ref{f2} (a),
a BAL trough is present at about 7000 \kmps\ with respect to the \heiozetz\ in the NIR spectrum.

\section{Emission Lines Analysis}
\subsection{UV \& Optical \feii\ Multiples }
\begin{figure*}[htbp]
\centering
\epsscale{0.6}
\plotone{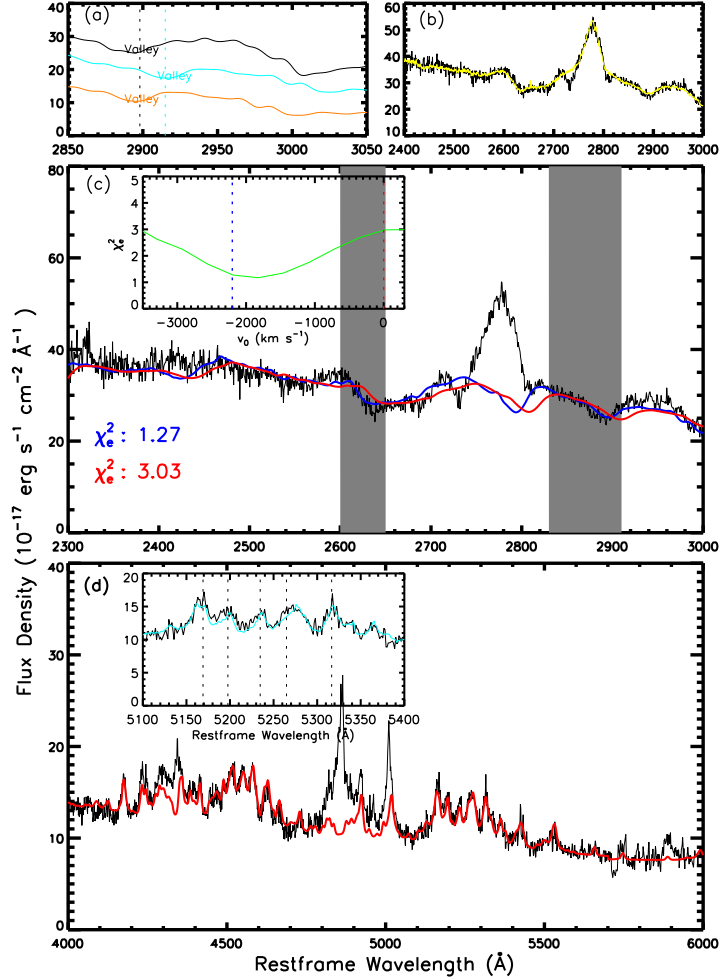}
\caption{ \textbf{Panel (\textit{a})}: The comparison between SDSS J1633+5137 (black) and scaled spectrum of
IZW1 (cyan) in the wavelength range from 2850\AA\ to 3050\AA. The black and cyan dashed lines mark the
valley between UV 60 and UV 61 spikes of SDSS J1633+5137 and IZW1, respectively. Vertical offsets have been
applied for clarity. After manually shifting the spectrum by --2200 \kmps\, the valley in the scaled spectrum of
IZW1 (orange) looks the same as that of SDSS J1633+5137.  \textbf{Panel (\textit{b})}:
The composite spectrum of five normal quasars (yellow) for which the UV \feii\ and \mgii\ are matched to that of
SDSS J1633+5137 (black). \textbf{Panel (\textit{c})}: The UV \feii\ fitting results for the blueshifted
velocity $v_0$ = -2200 \kmps\ (blue) and $v_0$ = 0 (red).
The corresponding reduced $\chi_e^2$ in the wavelength range marked with grey are also presented.
Besides, the variations of $\chi_e^2$ as a function of the blueshifted velocity of UV \feii\ are displayed.
Compared to $v_0=0$, the case of $v_0=2200$ \kmps\ is more acceptable for the fitting results.
\textbf{Panel (\textit{d})}: The fitting results (red) of optical \feii\ multiples with the blueshifted velocity fixed
at zero. The strong \feii\ lines (\feii\ 5169.03 \AA, 5197.57 \AA, 5234.62 \AA, 5264.8 \AA, and 5316.61 \AA) in the wavelength range from 5100 \AA\ to 5400 \AA\ are also marked by dotted lines and correspond to the peak of SDSS J1633+5137 spectrum. }
\label{f3}
\end{figure*}

   The \mgii\ broad emission line, which is dominated by the blueshifted component, is the most
   remarkable characteristic of SDSS J1633+5137. The blueshift velocity of \mgii\ peak is about 2200 \kmps.
   To precisely obtain the profile of \mgii\ emission line, the UV \feii\ multiples should be fitted and
   subtracted first. Interestingly, in the analysis of the UV \feii\ multiples, we find that they are
   also blueshifted and the blueshifted velocity is close to that of \mgii.
   This is supported by the following three evidences:

   First, the valley between the two spikes of \feii\ multiple UV 60 and UV 61 is an important feature
   in the UV \feii\ pseudocontinuum emission around \mgii.
   In the panel (a) of Fig.\ref{f3}, we present this valley in the SDSS J1633+5137 rest-frame in black curve.
   For comparison, we also plot the scaled spectrum of NLS1 AGN IZW1 in cyan (shifted to its
   rest-frame).
   Despite of the influence of \fei\ and \heitnff\ in 2930-3000\AA, the valley of SDSS J1633+5137
   shows blueshifted with respect to that of IZW1. To make it more clear, we manually blueshifted the scaled spectrum of
IZW1 by 2200 \kmps\ which is displayed by orange curve.
The valley seems to be consistent with that of SDSS J1633+5137, indicating that
the UV \feii\ multiples of SDSS J1633+5137 are blueshifted.

  Second, we searched for five normal quasars in the BOSS DR10, for which features of UV \feii\ multiples
 and \mgii\ are not blueshifted.
 After scaled with a power law curve, we constructed the composite spectrum of the five quasars and then
 matched it to the \mgii\ peak of SDSS J1633+5137, which is plotted in yellow line in the panel (b) of Fig.\ref{f3}.
 Since the UV \feii\ and \mgii\ in normal quasars are considered to have the same relative velocity with respect to
 the systemic redshift, the agreement of the normal quasars and SDSS J1633+5137 suggests that
  the blueshifted velocity of UV \feii\ multiples in SDSS J1633+5137 is close to that of \mgii.

  The last evidence comes from quantitative measurements of the UV \feii\ multiples. We used a combination of a single power law continuum and UV \feii\ multiples to fit the spectrum of SDSS J1633+5137 in the rest-frame wavelength range of  2200-3000 \AA. The model used to fit the observed spectrum can be described as
  \begin{equation}
 Model_{UV \feii}=C_{1} \lambda^{C_{2}}+C_3 f(v_0,\sigma).
 \end{equation}
 The $C_{1} \lambda^{C_{2}}$ is a power law which is used to fit the continuum and the $C_3 f(v_0,\sigma)$ is used to fit the UV \feii\ multiples. $v_0$  and $\sigma$ represents the shifted velocity and broadened width
 of UV \feii\ respectively. In the fitting process, the $v_0$ is fixed at a given value,
 the $C_1$, $C_2$, $C_3$ and $\sigma$ are free parameters and their best-fit values are searched by minimizing $\chi^2$.
 To distinguish the fitting results between different $v_0$ values given, we select the most remarkable
 UV \feii\ multiple, the red shape of UV1 and the gap between UV 60 and UV61, which are marked in gray-shaded
 region in the panel (c) of Fig.\ref{f3}, to calculate the reduced $\chi_e^2$. The $v_0$ is first fixed at 0, which means the
 UV \feii\ has no shift compared to the quasar's rest-frame. The result is displayed in red in the panel (c) of Fig.\ref{f3} with
 the reduced $\chi_e^2=3.03$. Then we fixed the $v_0$ at -2200 \kmps, which means that the UV \feii\ multiple is blueshifted at
 a velocity the same as that of \mgii.
The results is also displayed in blue with the reduced $\chi_e^2=1.27$, which suggests an obvious improvement
compared to $v_0=0$.
To display the variation of reduced $\chi_e^2$ as a function of $v_0$, we run a series of fitting programs where
a grid of $v_0$ is provided.
The $\chi_e^2$ variation with $v_0$ is plotted in the inset panel of Fig.\ref{f3}(c).
It can be seen that the reduced $\chi_e^2$ at $v_0=-2200$ \kmps\ is very close to the minimum value of reduced $\chi_e^2$ (1.17),
suggesting that the shift velocity of UV \feii\ is indeed close to that of \mgii.

   Previous studies for the UV \feii\ and optical \feii\ have shown that there is no obvious redshift offset between
   the two components (Sameshima et al. 2011). However, this conclusion is based on the quasar samples for which
   UV \feii\ and optical \feii\ are nearly at the systematic redshift. As mentioned above, the UV \feii\ multiples
   of SDSS J1633+5137 are supposed to be blueshifted with a velocity of about 2200 \kmps\ .
   It is not clear whether the optical \feii\ have the same blueshifted velocity in SDSS J1633+5137.
   Thus, we first compared the optical \feii\ of SDSS J1633+5137 (black) to the scaled spectrum of IZW1 (cyan)
   in the wavelength range of 5100\AA\ to 5400\AA, which is shown in the inset panel of Fig.\ref{f3} (d).
   Different from the UV \feii\ multiples, the peaks of strong \feii\ lines are close to those of IZW1, for which
   the shift velocity is corresponding to the source systematic redshift.
   Furthermore, the model with a single power law continuum and optical \feii\ multiples was also used to fit the
   spectrum of SDSS J1633+5137 in the wavelength range of 4000-6000\AA.
   The fitting results with the shift velocity fixed at 0 is plotted in red in Fig.\ref{f3} (d).
   Consistent with above empirical analysis, no obvious velocity shift is found.

   According to the analysis above, the UV \feii\ and optical \feii\ have different velocity shift.
   A reasonable assumption is that there are two \feii\ emitters excited in SDSS J1633+5137.
   One emits the blueshifted, strong UV \feii\ but faint optical \feii\,
   which could be arisen from the outflow gas (Gaskell 1982; Marziani et al. 1996; Leighly 2004).
   The another is from the normal BLR, where the UV \feii\ is faint but optical \feii\ is strong.
   Thus, we decomposed the UV-optical \feii\ multiples in SDSS J1633+5137 into two different components.
   One component is blueshifted and the another is at the quasar's systematic redshift.
   For each component, we employed the same program to fit the UV \feii\ and optical \feii\ as above.

\begin{figure*}[ht]
\epsscale{1.1}
\plotone{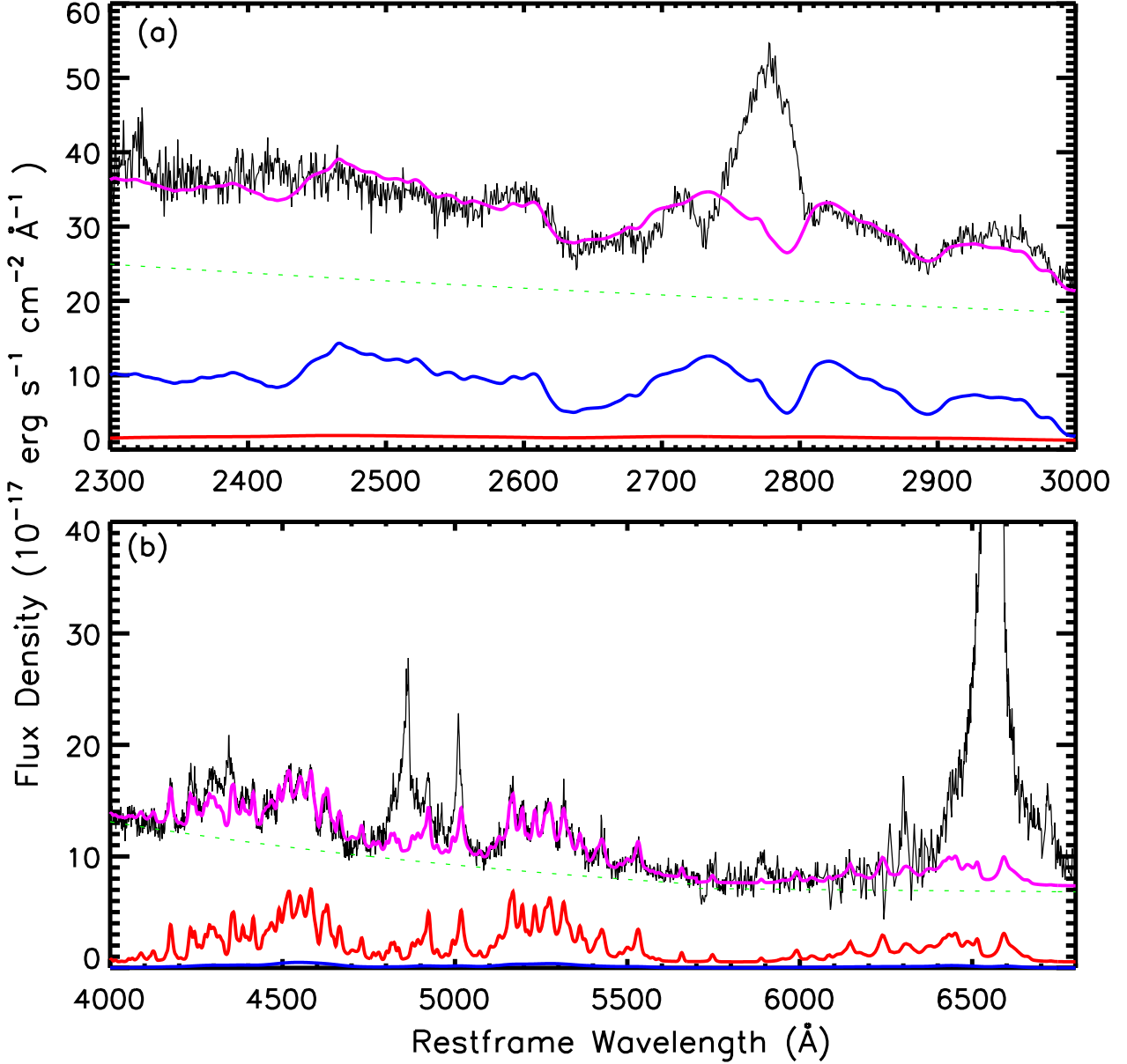}
\caption{The fitting results of UV and optical \feii\ multiples. The continuum is plotted in green.
The blueshifted and rest \feii\ components are displayed by blue and red curve respectively.
The sum of continuum is present in magenta. The best-fitted velocity of the blueshifted \feii\ component relative
to the quasar's rest-frame is -2135 \kmps ($\sim$-2200 \kmps) and that of rest \feii\ component is -92 \kmps ($\sim$0 \kmps).
Compared to the observed spectrum, the trough near 2730 \AA\ is considered to be \mgii\ absorption line.
The excess in 2930-3000\AA\ might be \fei\ and \heitnff.
The disagreement near 4300\AA\ is due to \hr\ and \oiiiftft\ emission lines.
}
\label{f4}
\end{figure*}

   In the fitting program, the shift velocity of UV-optical \feii\ is tied and allowed to vary.
   The first input shift velocity of the blueshifted component is -2200 \kmps\ and that of
   the rest component is 0.
   The fitting results of two components can be seen in Table 2 and shown in Fig.\ref{f4}.
   Similar to Sameshima et al.(2011), the total flux of optical \feii\ flux in 4435-4685 \AA\ is chosen
   as the intensity parameter of optical \feii\ multiples.
   The total flux of \feii\ UV1 multiples (2565-2665 \AA, Baldwin et al. 2004) is selected
   as the intensity parameter of UV \feii.
   It should be noted that the F-test results for the blueshifted optical \feii\ and rest UV \feii\ suggest
   that the two components are not required statistically during the fittings.
   Thus, we consider the fitting results of blueshifted optical \feii\ and rest UV \feii\ as their upper limits.

\subsection{Narrow Emission Lines }

   After subtracting the UV \& optical continuum and  UV \& optical \feii\ multiples, we are able to
   obtain the \mgii,  \hb\ broad emission line which is blended with \hb\ and \oiii\ narrow lines,
   and \ha\ broad emission line which is blended with \ha, \nii, and \sii\ narrow emission lines.
   With the help of the individual narrow emission line, \oii, we can derive the profiles of other
   narrow emission lines, which are then used to deblend the \hb\ and \ha\ broad lines.

\begin{figure*}[ht]
\epsscale{0.7}
\plotone{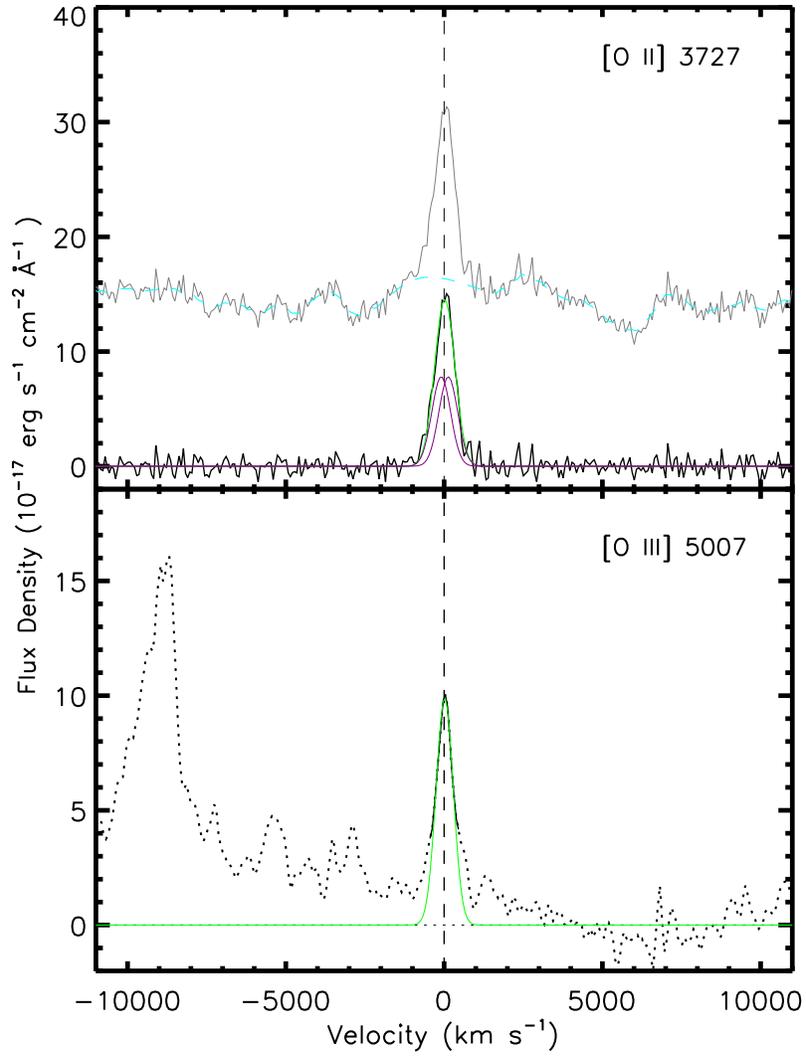}
\caption{The NELs of \oii\ and \oiii\ in SDSS J1633+5137. The two lines are fitted freely.
The modelled velocity shifts and widths of these NELs are very close.
}
\label{f5}
\end{figure*}

   To measure the \oii\ emission line, we masked out the spectrum in the velocity range -1500 to 1500 \kmps\ and used
   a 3-order spline curve to fit the local continuum of \oii. The local continuum is displayed by cyan dashed line
   in the top panel of Fig.\ref{f5}. The \oii\ includes two narrow emission lines, \oii\ 3729 and \oii\ 3726.
   For each narrow emission line, we used one Gaussian to fit its profile.
   The two Gaussians have the same profile in its own velocity space.
   The line ratio of \oii\ I(3729)/I(3726) is first fixed at 1.
   The fitting result is given in Table 2.
   According to Pardhan et al. (2006), the line ratio of \oii\ can vary from 0.35 to 1.5.
   We also try to model with different line ratios, while the width and wavelength shift are constrained to vary less than 30\kmps.
   If assuming the width of \oii\ is approximate to the velocity dispersion of the host bulge,
   the mass of the central BH log $M_{BH}/M_\odot$ can be estimated as 8.8 $\pm$ 1.3,
   according to the  Ferrarese \& Merritt (2000).

Besides \oii, we also tried to fit the \oiii\ 5007 narrow emission line despite it is blended with \hb\ broad emission line.
As shown in the bottom panel of Fig.\ref{f5}, the intensity of \hb\ BEL extended to the \oiii\ wavelength region is about $\sim$1
while the flux of \oiii\ NEL is $\sim$10. Thus, we conclude that the influence of \hb\ BEL can be ignored in the \oiii\ fittings.
We modeled the \oiii\ line with one Gaussian and the results are shown in Table 2.
Note that the modelled profile of \oiii\ is very close to that of \oii.

\subsection{Broad Emission Lines}
\begin{figure*}[ht]
\epsscale{0.7}
\plotone{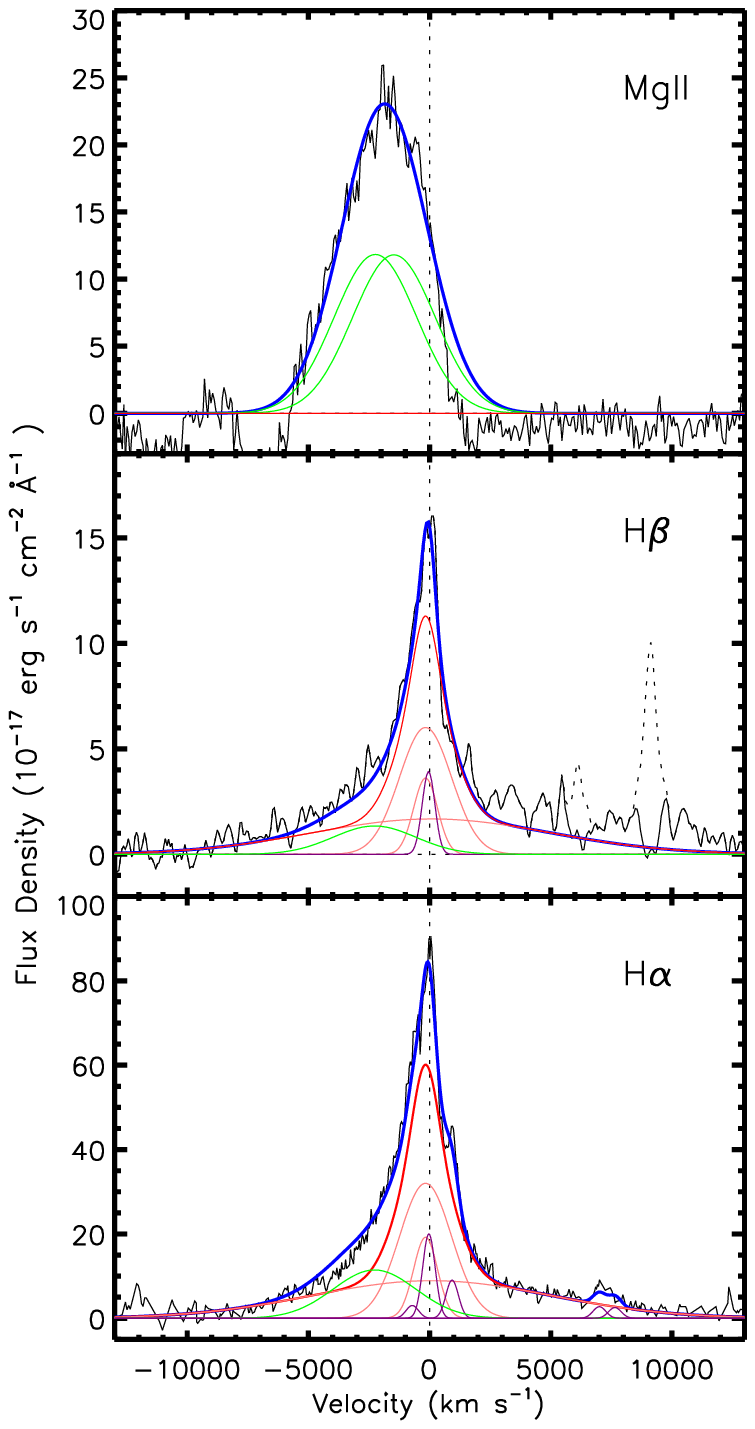}
\caption{The \mgii, \hb\ and \ha\ of SDSS J1633+5137 shown in their common velocity space.
From top to bottom, emission lines are sorted from shorter to longer wavelengths.
{The modelled \oiii\ (dotted) has been subtracted and the blueshifted \mgii\ doublets are present in green.}
The Balmer emission lines are decomposed into the broad (red)
and blueshifted (green) component.
The broad components of Balmer lines are fitted with three Gaussians,
each of which is present with pink curve.
The narrow emission lines are plotted in purple.  }
\label{f6}
\end{figure*}

    Based on the analysis above, we have derived the \mgii, \hb\ and \ha\ BELs in SDSS J1633+5137
    and these BELs are displayed in corresponding velocity space in Fig.\ref{f6}.
   Same as Wang et al. (2011), for a specific emission line, the parameter BAI is defined as the flux ratio of the blue part to the total profile, where the blue part is the portion of the emission line at wavelength less than
   its laboratory rest-frame wavelength. For \mgii\ doublet, the rest-frame wavelength is set to be 2999.4, which is obtained from the \mgii\ line core of IZW1. Based on this definition, we calculated the BAI of \mgii\ in our source and the value is $0.85 \pm 0.01$. We note that
   if we consider the possible existence of \mgii\ NEL, this value would be larger.
    It indicates that the \mgii\ is dominated by the blueshifted component.
    Similar to \mgii, the BAI of \hb\ and \ha\ is $\sim$0.56 and 0.54 respectively,
    suggesting that the blueshifted components of \hb\ and \ha\ BELs are also detected.

    Thus, we tried to decompose the \mgii, \hb, and \ha\ BELs into two components: one is blueshifted and emitted
    from the outflow, the another is in the quasar's rest-frame from the normal BLR.
    The blueshifted component was modelled with one Gaussian,
    while the component from normal BLR was fitted with multiple Gaussians.
    For the latter, we started from one Gaussian, and inspected visually the resulting
    $\chi^2$ and residuals to determine the goodness of fit.
    When the best possible fit was not achieved, we added another Gaussian with a relative
    velocity shift less than 100 \kmps. The fit was repeated until the $\chi^2$ was minimized
    with no further improvement in statistics.
    In the fitting progress, the intensity of each line is free except the ratio of \mgii\ doublets
    which was held fixed at 1:1.
    For SDSS J1633+5137, three Gaussians are good enough to fit the non-blueshifted BEL component.
    All these Gaussians were simultaneously fitted through the above iterative $\chi^2$-minimization process,
    and the fitting results are summarized in Table 2.
    Besides the BELs, the \hb\ and \ha\ also include
    the \nii, \sii, and Balmer NELs.
    Each NEL was modelled with one Gaussian, for which velocity shift and
    width were fixed at the values derived from \oii, assuming that all NELs in
    the spectrum have a similar profile to \oii.

    In the fitting progress, we noted that absorption troughs are present
    around \mgii\ emission lines. To further eliminate the effect of absorption lines,
    we first fitted the \mgii\ emission line with one Gaussian, and then masked our those pixels
    of absorption features deviating strongly from the model.
    In addition to NEL, \oiii\ always contains the blue
    outlier (e.g. Komossa et al. 2008; Zhang et al. 2011). For SDSS J1633+5137,
    however, the F-test suggests that another Gaussian for the blue outlier is not required in
    SDSS J1633+5137.
    Based on the profile of \hb\ rest component, we derived the mass of central BH of
    log $M_{BH}/M_\odot=8.37\pm$ 0.27, which is consistent with the mass estimated from \oii.

  The intensity ratio of blueshifted \mgii\ to \ha\ is useful to constrain the properties of outflowing gas.
    However, the decomposition of blueshifted \ha\ may be model-dependent, leading to uncertainty in the intensity ratio.
    We tried to determine its upper and lower limits. For the line ratio of \mgii\ to \ha, the lower limit
    can be estimated as shown in the left panel of Fig. 7.
    In this figure, the flux of H$\alpha$ and \mgii\ is normalized by the peak of H$\alpha$.
    The total \mgii\ emission line (red) is obviously
    blueshifted and its red side can reach about 1000 \kmps.
    Under the assumption that the rest component in the broad emission lines arises from the normal BLR for which
    the predominant motion is either Keplerian or virial
    (see Gaskell 2009 for a review), the \mgii\ rest component is expected symmetric.
    However, the red side of the observed \mgii\ is affected by the absorption line (Figure 6), and
    the red side of the modelled total \mgii\ reaches 3000 \kmps,
    This gives the blue side of rest component of -3000 \kmps for the rest component of \mgii.
    Thus, we selected the part of \mgii\ with relative velocity between -5000 \kmps and -3000 \kmps
    where the \mgii\ flux is prominent and the influence of \mgii\ rest component is small.
    For the \ha\ in the same relative velocity range, however, the emission line flux includes that
    of rest component.
    Hence the line ratio of blueshifted \mgii\ to \ha\ in this velocity range can be considered as
    the lower limit, which is estimated to be 0.46.
    As shown in the right panel of Fig. 7, based on the same assumption that the \ha\ rest component is symmetric, the lower limit of blueshifted broad \ha\ can be estimated by subtracting the symmetric flux on the blue side from the total.
The residual flux at the blue side is shown in green.
The line ratio of \mgii\ to \ha\ in the velocity range between -5000 \kmps\ and -3000 \kmps\
is close to 1, which can be considered as its upper limit.

 \subsection{Ionization Model for Blueshifted Emission Lines}

  Because the blueshifted velocities of UV \feii, \mgii, and Balmer lines are nearly the same,
  we supposed that these blueshifted components arise from the same outflowing gas.
  Thus, we can infer the properties of the outflows from these line ratios, using mainly the
  blueshifted \mgii/\ha\ and UV \feii/\ha.
  We did not use the \hb/\ha\, as the \hb\ and \ha\ blueshifted components are relatively weak in the emission lines.
  The line ratio between them may have large errors and hence be not reliable.
  For the blueshifted \mgii/\ha, as we discussed above, it was estimated to be $\sim$0.46--1.
  The blueshifted UV \feii\ to \ha\ is equal to the UV \feii/\mgii\ times \mgii/\ha.
  Because the blueshifted \ha\ component is relatively weak compared to total flux in the emission line,
  we expect that the error of \ha\ blueshifted component is much higher than the \mgii\ and UV \feii.
  Taking this into account, the error of blueshifted UV \feii/\ha\ mainly comes from the error of
  \mgii/\ha\ and the error of UV \feii/\mgii\ can be neglected.
  As shown in Table 2, the value of blueshifted UV \feii/\mgii\ is 0.75. Multiplied by the range
  of blueshifted \mgii/\ha, the blueshifted UV \feii/\ha\ can be estimated in the range $\sim$0.35-0.75.
  This is consistent with the \mgii/\ha\ (0.48) derived from the quasar composite spectrum (Vanden Berk et al. 2001).
  In spite of the blueshifted UV \feii/\ha\ is much larger than that from the quasar composite spectrum (0.01),
  it is consistent with the line ratio of a typical BLR derived from the photoionization model
  (Baldwin et al. 2004; Sameshima et al. 2011).
  Therefore, the UV \feii\ blueshifted components can be modelled with
  photoionization model and the physical conditions of outflowing gas are supposed to be
  similar to the BLR in normal quasars.

   The large-scale synthesis code CLOUDY (c13.03; Ferland et al. 1998) is employed to
   perform the photonization modeling of the blueshifted broad emission lines.
   The simulation results are used to compare with the luminosities and ratios of the
   blueshifted components measured from the spectrum of SDSS J1633+5137.
   In photoionization simulations, solar elemental abundance is adopted and the gas is
   assumed free of dust. To model the \feii\ emission lines, we used a 371 level $\rm Fe^+$ model
   that includes all energy levels up to 11.6 eV, and calculated strengths for 68,000 emission
   lines (Verner et al. 1999). For simplicity in the computation, the geometry is assumed as a slab-shaped emission medium with a uniform density, metallicity, and abundance.
   This medium is exposed to the ionizing continuum from the central engine
   with a SED defined by Mathews \& Ferland (1987, hereafter MF87).
   As shown in Fig. 8, an array of Hydrogen absorption column densities (N$_{\rm H}$)
   was set in the simulations
   from $10^{21}$ to $10^{24}$ cm$^{-2}$ stepped with 1 dex.
   As we discussed above, the ionization conditions may be nearly same as the BLR.
   Thus, for each column density, the range of outflow gas electron density (n$\rm _H$) was set
   from $10^7$ to $10^{14}$ cm$^{-3}$ and a grid of models are calculated by
   varying the $n_H$ of the emitting gas with a step of 0.5 dex.
   Finally, the logarithmic ionization parameter (log U) was sampled from -3.5 to 1.5 with a step of 0.5 dex.

  The calculated results are shown in Fig.8, where we plot the contours of blueshifted \mgii/\ha\ and
  UV \feii/\ha\ as a function of n$\rm _H$ and U. In each panel, the solid lines denote the basic models
  and the filled areas represent the observed range with $\rm 1 \sigma $ confidence level.
  The simulation results indicate that the observed regions of \mgii/\ha\ and UV \feii/\ha\
  have no overlap when N$\rm _H \leq 10^{22}$ cm$^{-2}$. The overlap region starts to appear
  for the column density N$\rm _H \geq 10^{23}$ cm$^{-2}$. Hence we considered  $10^{23}$ cm$^{-2}$
  as the lower limit on the column density, which implies that the outflow in SDSS J1633+5137 may be
  optically thick.

  Thus, we provided an ionization boundary model (Ferland et al. 1998) to simulate the emitting gas in the outflow
  and the simulation results are plotted in Fig. 9.

  In this model, the parameters n$\rm _H$ and U of the emitting gas can be constrained as of n$\rm _H$ from $10^{10.6}$ to $10^{11.3}$ cm$^{-3}$ and log U from -2.1 to -1.5. With n$\rm _H$ and U, the distance of emitting gas to central ionizing
  source was derived as $R_{emit} ~=~ (Q(H)/(4 \pi cUn_e) )^{0.5}$, where Q(H) is the number of ionizing photons,
  $Q(H) = \int_{\nu}^{\infty}{L_{\nu}/h \nu d \nu }$. Based on the continuum luminosity at 5100\AA ($\rm \lambda L_{\lambda} (5100\AA)~=~1.3 \times 10^{45} ~ erg~ s^{-1}$) and MF87 SED,
  we derived $\rm Q(H) \approx 1.1 \times 10^{56}~ photon~ s^{-1} $.
  Thus, we obtained that the distance between the emitting gas and central ionizing source is $\sim$0.1pc.
  Based on the best constrained parameter values by our photoionization modeling, namely,
  $\rm n_H$ = $10^{11.1}$ cm$^{-3}$ and log U = -1.8, we obtained the simulated EW of \mgii\ of 257 \AA.
  Since this value is modelled under the assumption of full sky coverage of outflowing gas,
  the ratio of observed EW of \mgii\ to the modelled one can be used to constrain the covering
  factor $C_{f,emit}$ of the emitting gas. The observed EW of \mgii\ is about 45 \AA, suggesting that
  the $C_{f,emit}$ is about 0.16.

    \section{Absorption Lines Analysis}
\subsection{Absorption-free Spectrum for the Absorption Lines}

As shown in Fig.\ref{f1}, a prominent BAL trough is present in the spectrum at about 7000 \kmps\ in velocity space
blueshifted with respect to the \heiozetz.

This trough can be identified as \heiozetz\ BAL.
Hinted by the location of the trough, we detected another BAL trough at about 7000 \kmps\ blueshifted
with respect to the \heiteen. In addition, at the same location in the respective velocity space,
the \mgii\ BAL was found in the spectrum. With these BALs, we are able to place constrains on the properties
of the absorption line outflowing gas.

To measure these BALs, we first used the pair-match method (Zhang et al. 2014; Liu et al. 2015) to
recover the absorption-free spectrum of SDSS J1633+5137. The absorption lines of interest
in the observed spectral regime include \heiozetz, \heiteen, \heitoen\ and \mgii. For each absorption, the pair-match method was employed to obtain the absorption-free spectrum.

 (1) \heiozetz\ regime: As can be seen from Fig.\ref{f10}, with a large blueshift of 7000 \kmps, the \heiozetz\ BAL is well detached from the corresponding emission line. This spectral regime is largely free from other emission lines (see the quasar composite spectrum displayed in Fig.\ref{f2}; Zhou et al. 2010). The absorption-free flux recovered by the pair-matching method is mostly contributed by the featureless continuum, which is well reproduced by a power-law and a black body emission.  We did not detect starlight from the host galaxy, and interpret the power-law component to be originated from the accretion disk of the quasar. The black body component is generally believed to be hot dust reradiation of the torus presumed by the AGN unification schemes (e.g., Netzer 1995). After removal of the black body component, we found that the residual flux is still significant in the BAL trough.
 (2) \heiteen\ regime: The absorption free flux around \heiteen\ BAL is mainly contributed by the power-law continuum radiated by the accretion disk. The absorption depth of the deepest part in the \heiteen\ BAL trough is $\sim 20\%$ on the normalized spectrum (see Fig.\ref{f10}). Since the absorption strength ratio ($gf_{ik}\lambda$) of  \heiozetz\ to \heiteen\ is as large as 23.3 (e.g., Leighly et al. 2011), this indicates that the BAL region only partially covers the accretion disk, incorporating the fact that there is still significant residuals in the  \heiozetz\  trough after removal of the host dust contribution. Detailed analysis yielded the covering factor of the absorption gas to the accretion disk is about 0.4 (see Sec 4.2)

 {(3) \heitoen\ and \mgii\ regime:
 The emission and absorption characteristics around these two absorption lines are nearly the same. }
 The power law continuum from the accretion disk and the \mgii\ and \feii\ broad lines from the outflow.
 Considering that the absorption gas only partially covers the accretion disk, and the emission line gas
 is of the similar size as the normal BLR, the UV \feii\ multiples should not be included in the absorption free
 spectrum. The normalized absorption spectra of \heiozetz, \heiteen, \heitoen, and \mgii\ are displayed in Fig.\ref{f10} (right).

 \subsection{Characterizing the Absorption Line Gas}
    Before investigating the properties of BALs, we first constrain the distance of BAL outflow gas
    in a qualitative way. According to the discussion above, the absorption medium partially obscures the
    accretion disk. Thus, we considered that the distance of the absorption medium is comparable to the
    size of accretion disk at 10830 \AA. Based on the equation 3.2 in Peterson (1997), the size of accretion
    disk at 10830 \AA\ is about 1500 $r_g$ ($r_{g} \equiv G{M_{BH}}/c^2 $), or 0.017pc.
     More quantitive constraint on the distance of the
      absorption gas requires the measurements of ionization
      parameter U and gas density $n_H$. The latter can be constrained
      by comparing the photoionization simulations with observed line
      ratios of multiple ions (e.g., Leighly et al. 2011; Liu et
      al. 2016). However,  only He I$^*$ and \mgii\ absorptions are
detected in J1633. As we will show below, they are not sufficient to set useful constraint on
the gas density, but useful in determining the ionization parameter and  lower limit on the total column density for the BAL gas.  


 For a BAL, the normalized intensity is
 \begin{equation}
 I(v) = [1-C_f(v)]+C_f(v)e^{-\tau(v)},
 \end{equation}
  where $C_f(v)$ is the covering factor and $\tau$ is true optical depth as a function of
 radial velocity. For transitions from the ion at a given level, the values of $\tau(v)$
 is proportional to f$\lambda N_{col}$, where f is the oscillator strength,  $\lambda$ is the rest wavelength of the transition and $N_{col}$ is the column density of the ion with the given level.
 Theoretically, two absorption lines transited from the same ion at a given level are
 needed to derive the physical conditions of outflowing gas, such as $C_f$ and $N_{col}$.

 For  SDSS J1633+5137, three absorption lines, \heitoen, \heiteen\ and \heiozetz\ are transited from the same
 energy level \hei\ and can be used to derive the $C_f$ and $N_{col}$ of this ion.
 As the \heitoen\ trough is weak, we used the \heiteen\ and \heiozetz\ to solve the equation (2) to
 obtain \cf\ and \tauv\ of HeI*.
 The \heitoen\ trough is employed to check for consistency.
 Because the \heiozetz\ absorption line is affected seriously by the sky lines,
 the pixels in these region was marked and the data were interpolated using 3-order spline.
 As the bottom of \heiozetz\ is about 0.6, we defined that the edge of absorption trough region
 is located at the pixels where three continuous pixels are below 0.96,
 corresponding to a depth of absorption of 4\%, or 10\% of the depth of \heiozetz\ BAL.
 For every pixel in the absorption line regions of \heiteen\ and \heiozetz,
 we derived the $C_f$ and $N_{col}$ of \hei, and $\tau_{\rm HeI*3889}$, which are shown in Fig.\ref{f11}.

 The integral $N_{col}$ of He I$^*$ along with the absorption trough in the velocity space
 is found to be $(5.0 \pm 1.7) \times 10^{14}$ cm$^{-2}$.
 With the $C_f$ and $N_{col}$, we also simulated the absorption trough of \heitoen\ and compared it
 with the observed data in Fig.\ref{f12}. We found that the simulated and observed absorption
 trough are consistent with each other, indicating that the derived $C_f$ and $N_{col}$ are
 reliable.
 In addition, with the derived $C_f$ and \mgii\ absorption trough, we tried to constrain the
 $N_{col}$ of $Mg^+$.
 {In Fig.\ref{f13}, we show the trough of \mgii\ in blue and 1-$C_f$ in comparison,
 which indicates the saturation of \mgii\ absorption trough at some velocities.
 Thus, the $N_{col}$ of $Mg^+$ can not be obtained directly from Eq. 2.}
 However, because of the saturation of \mgii, we can derive log U from the $N_{col}$ of \hei\
 through the equation (3)
 in Ji et al. (2015) and the value of log U is -1.9 $\pm$ 0.2.
{\bf This value is in the range of e
 log U derived from the blueshifted emission lines.
}

 With our measurement for the total column density seen in the He I metastable lines, we
 can set a minimal He$^+$ column density of $\sim1\times10^{20}$ cm$^{-2}$ in the outflow,
 taking the maximum density ratio of HeI$^*$ to He$^+$ (Rudy et al. 1985; Arav et al. 2001).
 Assuming solar abundances this estimate yields a minimal H II column
 density N$_{\rm H}\sim1\times10^{21}$ cm$^{-2}$.
 On the other hand, we can estimate the HII column density of BAL gas through the equation
 $\rm N_{H} \approx 23+log U$ (Ji et al. 2015), yielding N$_{\rm H}\sim10^{21}$ cm$^{-2}$.
 However, it should be noted that the $N_{col}$ of \hei\ is not a suitable indicator of $\rm N_{H}$
 for the optical thick gas. This is because HeI* is a high-ionization line and its column
 density mainly grows in the very front of hydrogen ionization front and stops growing behind it
 (e.g., Arav et al. 2001; Ji et al. 2015). Instead, absorption lines with lower ionization
    potentials, such as CaII, \mgii, and \feii, are useful to probe the total column density
    of outflow. Unfortunately, the CaII and \feii absorption lines are not detected, and
    $N_{col}$ of $Mg+$ is difficult to derive due to the saturation effect.
Therefore, we can only set a lower limit for the total
HII column density of BAL gas, N$_{\rm H}>1\times10^{21}$ cm$^{-2}$.

On the other hand, further constraint on the column density of absorbing medium
can be placed with the non-detections of the
corresponding UV \feii\ BALs.
This is because given the same ionization parameter,
the MgII and FeII absorption lines are both sensitive to the total column density.
Similar to our analysis with BEL outflow (Section 3.4), we employ the photoionization simulations to
evaluate the dependence of FeII BALs on the total column density.
We assume the geometry of BEL gas as a slab-shaped medium exposed to the ionizing continuum from the central
engine with uniform density.
The model setups are the same as that for the BEL simulations except for the ionization parameter, which is
logU = -1.9 as derived from HeI* BALs.
Each individual simulation model is customized in terms of the column density ($\rm N_H$), which is
set to vary in the range 21 $\le$ log$\rm N_H$ $\rm (cm^{-2})$ $\le$ 22 with a dex step of 0.2.
This model can predict the population on various levels of Fe$ ^+ $ and the strength of absorption lines originated from these levels.
Fig.14 (upper panel) presents a series of models with the grid of log N$\rm _{H}$.
To better visualize the models, each photoionization model is broadened using the absorption profile of \hei~.
As can be seen from the simulation results, at log$\rm N_H$ $\ge$ 21.4, there are obvious absorption troughs from
the iron multiplets raised from the ground state (e.g.,
Fe II UV2+3 at approximately 2400 \AA\ and Fe II UV1 at approximately 2600 \AA. Due to the BALs are 7000 \kmps\ blueshifted, the ). Such absorption features are, however, not
observed in the spectrum. Thus the upper limit on the column density of BAL gas can be constrained to be
 at log$\rm N_H$ = 21.4. In fact, when compared to the observed spectrum in detail (Fig.14, lower panel), we found a model
  with column density of log$\rm N_H$ = 21.2 matches the data well in the
  spectral range of 2300-3000\AA.
Therefore, in combination with the lower limit on the column density
given by the
HeI* BAL,  the most probable column density for the BAL gas is log$\rm N_H$ $\sim$ 21.2. This suggests that
   the physical conditions of BAL and BEL gas are not strictly the same, at least in terms of the total column density.

\section{Summary and Discussion}
In this paper, we present a detailed study of the emission and absorption line properties of J1633+5137.
In the optical and NIR spectra, in addition to the normal emission lines
originating from BLR and NLR, there are several blueshifted emission components
with a common velocity at $\sim-2200$km s$^{-1}$ from MgII, UV FeII and hydrogen Balmer lines, suggestive
of the AGN BEL outflows.
These lines can be mutually used to constrain the physical properties of the outflowing gas
by confronting the observations with the photoionization simulations.
The physical parameters for the BEL outflow are constrained to be
$10^{10.6}$ $\le $ $\rm n_H$  $\le$  $10^{11.3}$ cm$^{-3}$, -2.1 $\le$ log$ U_E$ $\le$ -1.5,
and $\rm N_H$ $ \simgt $ $10^{23}$ cm$^{-2}$.
Using the ionization parameter, gas density and EW of MgII, we
estimated the covering factor and distance of the BEL outflow materials to the central source,
which is $C_{\rm f,emit}$ $\sim$ 0.16 and $r$ $\sim$ 0.1pc.
In addition, strong BALs from \mgii\ and HeI* metastable lines are also detected.
Using a simple partial coverage model, we derived the integral column density of HeI* and
ionization parameter for the BAL gas, ,
which is $(5.0\pm1.7)\times10^{14}$ cm$^{-2}$ and logU = -1.9 $\pm$ 0.2, respectively.
The total column density is estimated in the range  $10^{21}$ $\le$ log(N$_{\rm H})$ $\le$ $10^{21.4}$ cm$^{-2}$,
which is about two orders of magnitude less than that derived for the BEL gas, suggesting that
the physical conditions of BAL and BEL gas are not strictly the same.

Though the blushifted BELs are crucial in studying AGN outflows which can reflect the
global properties of outflowing gas, their physical conditions and locations
are difficult to investigate, except for a limited number of sources where
the spectra from multiple ionic species can be reliably measured.
Liu et al. (2016) identified both the BELs and BALs produced by the
AGN outflows in the quasar SDSS J163459.82+204936.0.
The physical parameters determined for the BEL and BAL outflows are very
close, with $10^{4.5}$ $\le$ $\rm n_H$ $\le$ $10^{5}$ cm$^{-3}$, -1.3 $\le$ log$U$ $\le$ -1.0, $\rm N_H$ $ \sim$ $10^{22.5}$ cm$^{-2}$, and
the outflow materials are 48--65 pc from the central source, likely exterior of the
torus. The similarity of the physical parameters strongly suggest that
blueshifted-BEL and BALs should be generated in the common outflowing gas.
Zhang et al. (2017) reported similar UV and optical emission line outflows in the
heavily obscured quasar SDSS J000610.67+121501.2, and inferred a distance at the scale of
the dusty torus (and beyond).
Conversely, the emission line outflow identified in J1633+5137 has a much higher density ($\rm n_H$ $\sim$ $10^{11}$ cm$^{-3}$)
with a distance at the scale of BLR to the central source, reflecting the diversity
of physical conditions for the outflowing gas.

\subsection{Energetic Properties of the Outflow}

Since the physical conditions for the BELs and BALs are not the same,
we discuss separately the energetic properties of the BEL and BAL outflows.
As discussed in Borguet et al. (2012),
   assuming that the BEL outflowing material are described as a thin ($\Delta R/R \ll 1$ ),
   partially filled shell, the mass-outflow rate ($\dot{M}$) and kinetic luminosity ($\dot{E_{k}}$)
   are given by
 \begin{equation}
 \dot{M}=4 \pi R \Omega \mu m_{p} N_{H} v
 \label{fun:M}
 \end{equation}
 and
 \begin{equation}
 \dot{E_{k}}=2 \pi R \Omega \mu m_{p} N_{H} v^3
 \label{fun:EK}
 \end{equation}
 , where R is the distance of the outflow from the central source, $\Omega$ is the global covering
fraction of the outflow, $\mu = 1.4$ is the mean atomic mass per proton, $m_{p}$ is the mass of proton.
N$\rm _{H}$ is the total hydrogen column density of the outflow gas. $v$ is the radial velocity.
Based on the physical parameters inferred for the BEL outflow, and
 taking the velocity of outflow $v$ as the peak of blueshifted \mgii\ BEL, which is -2200 \kmps,
 the mass-outflow rate and the kinetic luminosity can be derived as $\dot{M}$ = 0.9 M$ _{\odot} $ yr$ ^{-1} $ and $\dot{E_{k}}$ = 1.5 $\times$ 10$ ^{42} $ erg s$ ^{-1} $, respectively.


Similar to the BEL outflow, we can also obtain the $ \dot{M} $ and $ \dot{E_k} $ for the BAL outflow.
However, the global covering factor and density of BAL outflow gas in SDSS J1633+5127 cannot be directly
constrained by the observations. In the studies of BAL quasars, the
global covering fraction of BAL outflow gas is generally derived from the fraction of BAL quasars. This fraction is
about 10\%-20\% in optical-selected quasars (e.g., Trump
et al. 2006; Gibson et al. 2010; Zhang et al. 2014).
Moreover, we assumed that BAL outflow locates at the same distance to the central source as the BEL outflow.
With the column density of BAL outflow log N$ \rm _{H} $ (cm$ ^{-2} $) = 21.2 and radial velocity of $\sim$7000 km s$^{-1}$,
the $ \dot{M} $  and $ \dot{E_k} $ for the BAL outflow can be estimated as $ \dot{M} $ = 0.01 M$ _{\odot} $ yr$ ^{-1} $ and $ \dot{E_k} $ = 2.2 $ \times $ 10$ ^{41} $ erg s$ ^{-1} $, respectively. These values are a factor of 7-9 less than
that obtained for the BEL outflow.
Therefore, the mass flux and kinetic luminosity are dominated by the BEL outflow,
and the contribution from BAL outflow is minor.


   Previous studies suggest that efficient AGN feedback in the form of high-velocity outflows typically requires kinetic luminosity
   to be the order of a few percent of the Eddington luminosity (L$ _{EDD} $)(e.g., Scannapieco \& Oh 2004; Di
Matteo et al. 2005; Hopkins \& Elvis 2010). For SDSS J1633+5127, the mass of black hole (log M$ _{BH} $/M$ _{\odot} $) derived from \hb\ is about 8.37 and L$ _{EDD} $ is about 3$ \times $ 10$^{46} $ erg s$ ^{-1} $. Taking the calculation results above,
the sum of kinetic luminosities of the BEL and BAL outflow is only $\sim$1.7 $ \times $ 10$^{42} $ erg s$ ^{-1} $ ( $ < $ 10$ ^{-4} $ L$ _{EDD} $). This value is apparently far from efficient to drive the AGN feedback.
Note that the kinetic luminosity of the total outflow gas can be considered only a lower limit for the following reasons:
(1) The column density  of 10$ ^{23} $ cm$ ^{-2} $ we inferred for the BEL outflow is the lower limit.
(2) The velocity $ v $ for BEL gas is a sum of the projected velocities of the outflowing gas
along different directions, the value of which is only a lower limit on the outflow velocity (Liu et al. 2016; Zhang et al. 2017).
(3) The distance of the BAL outflowing gas may also be a lower limit, as if it was located at much greater distances from the central source.

\subsection{Outflow Geometry and the Profile of Outflow Emission Line}

As we mentioned above, blueshifted BELs from multiple ionic species are rarely observed in quasars, and
both of the BELs and BALs are observed in the spectrum of the same quasar is even rare.
In Section 3 and 4, we investigated the physical properties of the BEL and BAL outflows respectively,
and obtained similar ionization parameters for them.
Therefore, though the physical conditions are not strictly the same, the BEL and BAL outflows may not be independent.
In order to further constrain the outflow geometry,
we attempted to reproduce the profile of BELs with the radial velocity of BALs.


The outflows are always considered as a biconical structure in the previous works
 (e.g. Elvis 2000) and the emission line profile can be successfully modeled with this structure
 (Zheng, Binette, \& Sulentic 1990; Marziani et al. 1993; Sulentic et al. 1995).
 However, the biconical structure is two-dimensional and need a certain number of free parameters
 to reproduce the emission line profile in models. For simplicity,
 we employed a one-dimensional ``ring'' model to reproduce the emission line outflow profile of
 SDSS J1633+5137. The cross-section of this model is displayed in the left panel of Fig.\ref{f15}.

 The ring model for the outflow assumes that the line originates on a ring above the disk
 for which axis inclines with an angle $ i $ relative to the line of sight.
 The ring has an angle $ \theta_r $ relative to the normal direction of the accretion disk.
 For SDSS J1633+5137, as the blueshifted BELs and BALs of the outflow are observed at the same time,
 it is natural to assume our line of sight is penetrating through the outflow,
 which means the angel $i$ = $ \theta_r $.
 {To reproduce the velocity range of BALs, the $v_r$ in this model is constrained
 at 7000 \kmps, which corresponds to the blueshifted velocity of the BALs.}
 The distance from the ring to the black hole is $r$ (expressed in units of the gravitational
 radius, $r_{g}$).
 The distance of outflow to the central source derived from blueshifted BELs is about
 0.1pc and about 9000 $r_{g}$.
 The outflow velocity along with the radial direction is $v_r$.
 Besides the radial velocity, the outflow ring also has a rotation velocity.
 However, if we assumed our outflow is launched by the diskwind which arises from the accretion disk
 at about 100 $r_g$. When the outflow arrives at 9000 $r_g$, due to the angular momentum conservation,
 the rotation velocity is about 300 \kmps. This rotation velocity is much less than the $v_r$.
 Therefore, in our model, the rotation velocity was ignored. The coordinate of gas in the ring can be
 expressed as( $r, \theta_r, \phi $), where the $\phi$ changes from -$\pi$ to $\pi$ and our line of sight
 corresponds to $\phi$ = 0. Assuming the outflowing ring's rotation increases with the direction of $\phi$,
 for a certain ring at $r, \theta_r$, and $\phi $, the velocity on the line of sight $v_{obs}$ can be expressed
 as

   \begin{equation}
     v_{obs}(r,\theta_r, \phi)=-(v_rsin^2(\theta_r)cos(\phi)+v_rcos^2(\theta_r)).
   \label{functions:vobs}
   \end{equation}

To compare with the observed profile easily, we defined the direction of far away the central BH as
the positive direction of $v_{obs}$. With this equation, we can derive the emission line profiles of
the outflow ring for only one free parameter $\theta_r$ in our model.

In the right panel of Fig.\ref{f17}, we display three model results.
For comparison, we also show the fitting result to the \mgii\ blueshifted BEL (black).
All the three models are different from the \mgii\ blueshifted BEL.
The model profile is single-peaked when the $\theta_r$ is small,
but the blueshifted velocity is higher than \mgii.
For the case of larger $\theta_r$, the model profile becomes double-peaked,
which is also inconsistent with \mgii.
Even though, we found that when the $\theta_r$ = 40$^\circ$,
the red part of model profile appears to match well with the red side of \mgii.
If the emission at the blue side is obscured under a certain condition, and
only the emission at the red side can be observed,
the modelled profile could be consistent with \mgii.

According to the Eq.5, for a certain $\theta_r$, the larger blueshifted velocity corresponds
to the smaller absolute value of $\phi$. Thus, we proposed another amended toy model.
All the parameters in this model are the same to the model above except we added a free parameter
"shadow". The top view of this model is shown in the left panel of Fig.\ref{f16}.
The parameter "shadow" is in the range from 0 to 1.
For a given shadow parameter, the outflowing gas in the range $-shadow \times \pi$ to $shadow \times \pi$
is obscured and only the photons emitted from the rest of the ring can be detected.
In the middle panel of Fig.\ref{f16}, we display a modelled profile which can reproduce the
profile of \mgii\ well. The parameters of this best-fitted profile are
$\theta_r$ = 40$^\circ$ and shadow = 0.48.
Note that the free parameter $v_r$ and $shadow$ can be well constrained
in our model.
Fig.\ref{f16} (right panel) shows the 1, 2, and 3$\sigma$ confidence levels of the parameter $\theta_r$
versus $shadow$.
At 1$\sigma$ confidence level, the $\theta_r$ was constrained in the range from 33$^\circ$ to 43$^\circ$, while
the $shadow$ was from 0.33 to 1.
In Section 3.4, we have estimated that the distance of BEL gas is $\sim$0.1pc,
which is much smaller than the distance of dust torus, typically at $\sim$pc scale (e.g., Barvainis 1987; Koshida et al. 2014;
Kishimoto et al. 2012).
In addition, previous studies of the dust torus yielded the dust covering factors of $\sim$48$^\circ$
(Schmitt et al. 2001) and in a large range of 38-56$^\circ$ (Osterbrock \& Martel 1993; Sazonov et al.2015).
Therefore, it is possible that the shielding of outflowing gas in SDSS J1633+5137
is the dusty torus.
Future spectropolarimetry observations will be required to further test this model,
and helpful to place new constrains on the geometry of the outflowing gas.
It should be noted that this model is based on the assumption that the
density and column density are uniformly distributed in the outflow gas, which may be over-simplified.
Using more complex models, e.g., the column density distribution is inhomogenous, to explain the emission and absorption features observed in
SDSS J1633+5127 is beyond the scope of this paper. We present in the Appendix such a multi-column density modelling of
the outflow emission lines, and a detailed investigation will be presented elsewhere.

This work is supported by the National Natural Science
Foundation of China (NSFC-11573024, 11473025, 11421303, {11573001 and 11822301})
and the National Basic Research Program of China (the 973
Program 2013CB834905 {and 2015CB857005}). T.J. is supported by the National
Natural Science Foundation of China (NSFC-11503022) and
the Natural Science Foundation of Shanghai (NO. 15ZR1444200). P.J. is supported by the National Natural Science Foundation of China (NSFC-11233002).
{X.S. acknowledges support Anhui Provincial NSF (1608085QA06)
and Young Wanjiang Scholar program. }
We acknowledge the use of the Hale 200-inch Telescope at Palomar
Observatory through the Telescope Access Program (TAP), as
well as the archive data from the SDSS, 2MASS, and WISE
surveys. TAP is funded by the Strategic Priority Research
Program, the Emergence of Cosmological Structures
(XDB09000000), National Astronomical Observatories, Chinese Academy of Sciences, and the Special Fund for
Astronomy from the Ministry of Finance. Observations
obtained with the Hale Telescope at Palomar Observatory
were obtained as part of an agreement between the National
Astronomical Observatories, Chinese Academy of Sciences,
and the California Institute of Technology. Funding for SDSS-
III has been provided by the Alfred P. Sloan Foundation, the
Participating Institutions, the National Science Foundation, and
the U.S. Department of Energy Office of Science. The SDSS-
III Web site is http://www.sdss3.org/.

\begin{deluxetable}{cccc}
\tabletypesize{\scriptsize}
\tablewidth{0pt}
\tablenum{1}
\tablecaption{Photometric Observations of SDSS  J1633+5127}
\tablehead{
\colhead{Wavelength Band/Range}  & Mag. &  Survey  & \textit{MJD}  }
\startdata
\  $\textit{u}$          &$ 18.59\pm 0.04 $ &SDSS     &51948\\
\  $\textit{g}$          &$ 18.04\pm 0.01 $ &SDSS     &51948\\
\  $\textit{r}$          &$ 18.23\pm 0.01 $ &SDSS     &51948\\
\  $\textit{i}$          &$ 17.80\pm 0.01 $ &SDSS     &51948\\
\  $\textit{z}$          &$ 17.76\pm 0.02 $ &SDSS     &51948\\
$  J  $               &$ 16.57\pm 0.13 $ &2MASS   & 50937\\
$  H  $               &$ 15.73\pm 0.16 $ &2MASS   & 50937\\
$  K_{s}  $               &$ 14.41\pm 0.08 $ &2MASS   & 50937\\
$  W1 $               &$ 12.25\pm 0.02 $ &WISE   &55332 \\
$  W2 $               &$ 10.93\pm 0.02 $ &WISE   &55332 \\
$  W3 $               &$ 8.31\pm 0.02 $ &WISE   &55332 \\
$  W4 $               &$ 6.22\pm 0.04  $ &WISE   &55332 \\
$  V  $               &-                 &Catalina&53653-56454\\
\enddata

\end{deluxetable}

\begin{deluxetable}{lccc}
\tabletypesize{\scriptsize}
\tablewidth{0pt}
\tablenum{2}
\tablecaption{Decomposition Measurements of Emission Lines}
\tablehead{
\colhead{}& $\rm Int.^a$ & $\rm Shift^b$ & $\rm FWHM^b$}
\startdata
\oii\ 3729& $72 \pm 4$ & $21 \pm 4$     &$700 \pm 22$\\
\oiii\ 5008& $111 \pm 13$ & $26 \pm 5$     &$677 \pm 35$\\
blueshift \mgii\ 2796      &$457 \pm 4$       &$-2210 \pm 8$   &$3889 \pm 20$\\
blueshift \mgii\ 2803      &$457 \pm 4$       &$-2210 \pm 8$   &$3889 \pm 20$\\
rest \hb\        &$671 \pm 12$      &$39 \pm 18$     &$2623 \pm 71$ \\
rest \hb\ Gaussian1 &$336 \pm 6$      &$81 \pm 15$     &$11242 \pm 41$ \\
rest \hb\ Gaussian2 &$72 \pm 8$      &$-65 \pm 12$     &$1161 \pm 54$ \\
rest \hb\ Gaussian3 &$263 \pm 6$      &$-61 \pm 13$     &$2541 \pm 25$ \\
blueshift \hb\        &$93 \pm 6$       &$-2210 \pm 8$   &$3889 \pm 20$\\
rest \ha\        &$4836 \pm 47$     &$19 \pm 14$     &$2623 \pm 73$ \\
rest \ha\ Gaussian1 &$2420 \pm 43$      &$81 \pm 15$     &$11242 \pm 41$ \\
rest \ha\ Gaussian2 &$522 \pm 56$      &$-65 \pm 12$     &$1161 \pm 54$ \\
rest \hb\ Gaussian3 &$1892 \pm 46$      &$-61 \pm 13$     &$2541 \pm 25$ \\
blueshift \ha\        &$1069 \pm 40$     &$-2210 \pm 8$   &$3889 \pm 20$\\
$\textit{rest UV  \feii\ } ^{c,d}$  &$96 \pm 54$     &$-92 \pm 87$  &$8133 \pm 103$\\
blueshift UV $\rm \feii\ ^e$  &$623 \pm 47$      &$-2135 \pm 52$  &$1417 \pm 31$\\
rest optical $\rm \feii\ ^f$  &$1002 \pm 20$     &$ -92 \pm 87 $   &$2544 \pm 86$\\
$\textit{blueshift optical \feii\ } ^{c,g}$  &$189 \pm 19$      &$-2135 \pm 52$  &$8372 \pm 95$\\
\enddata
\tablecomments{\\
\textbf{a}:   In units of $\rm 10^{-17} erg~s^{-1}~cm^{-2}$.\\
\textbf{b}:   In units of \kmps.\\
\textbf{c}:   The upper limits.\\
\textbf{d}:   The total computed UV \feii\ flux over the wavelength ranges 2565-2665 \AA.\\
\textbf{e}:   The total computed blueshifted UV \feii\ flux over the wavelength ranges 2548-2647 \AA\ in the quasar's
rest-frame.\\
\textbf{f}:   The total computed optical \feii\ flux over the wavelength ranges 4435-4685 \AA.\\
\textbf{g}:    The total computed optical \feii\ flux over the wavelength ranges 4405-4653 \AA\ in the quasar's rest-frame.\\
}
\end{deluxetable}

\begin{figure*}[ht]
\epsscale{1.0}
\plotone{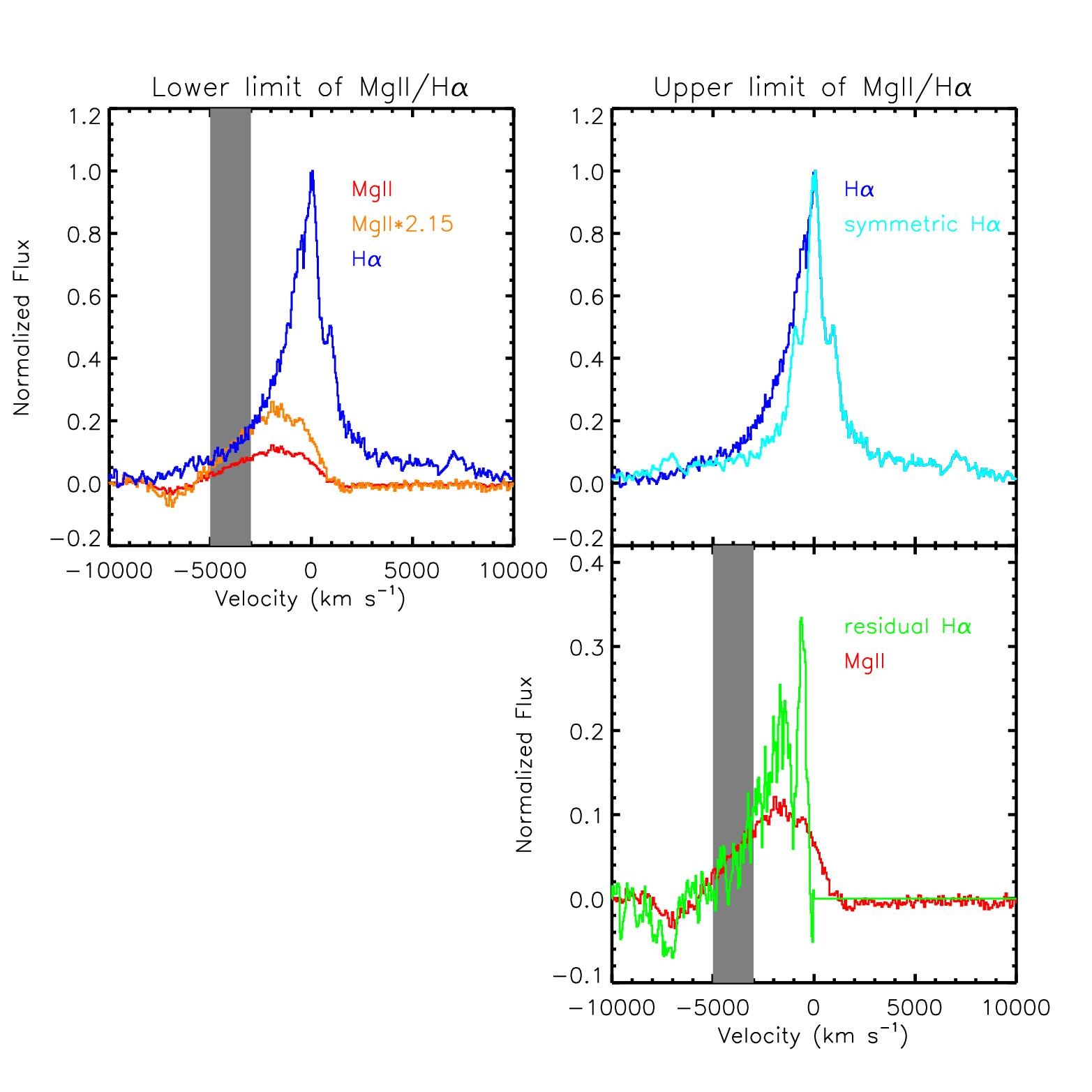}
\caption{ Estimation of lower and upper limit of  \mgii/\ha. The \mgii\ and \ha\ flux are all normalized by the peak of \ha. \textbf{Left}:
The \ha\ flux in the wavelength range between -5000 to -3000 \kmps\ comprises the photons
emitted from BLR hence the flux of blueshifted component may be overestimated, providing the lower limit of \mgii/\ha.
\textbf{Right}: The estimation for the upper limit of \mgii/\ha. The \ha\ in
3000-5000 \kmps\ may include the photons of blueshifted componet. Thus,
the rest component of mirror symmetric \ha\ flux from -5000 to -3000 \kmps\ may be underestimated,
giving the upper limit of \mgii/\ha.
}
\label{f7}
\end{figure*}

\begin{figure*}[ht]
\epsscale{0.8}
\plotone{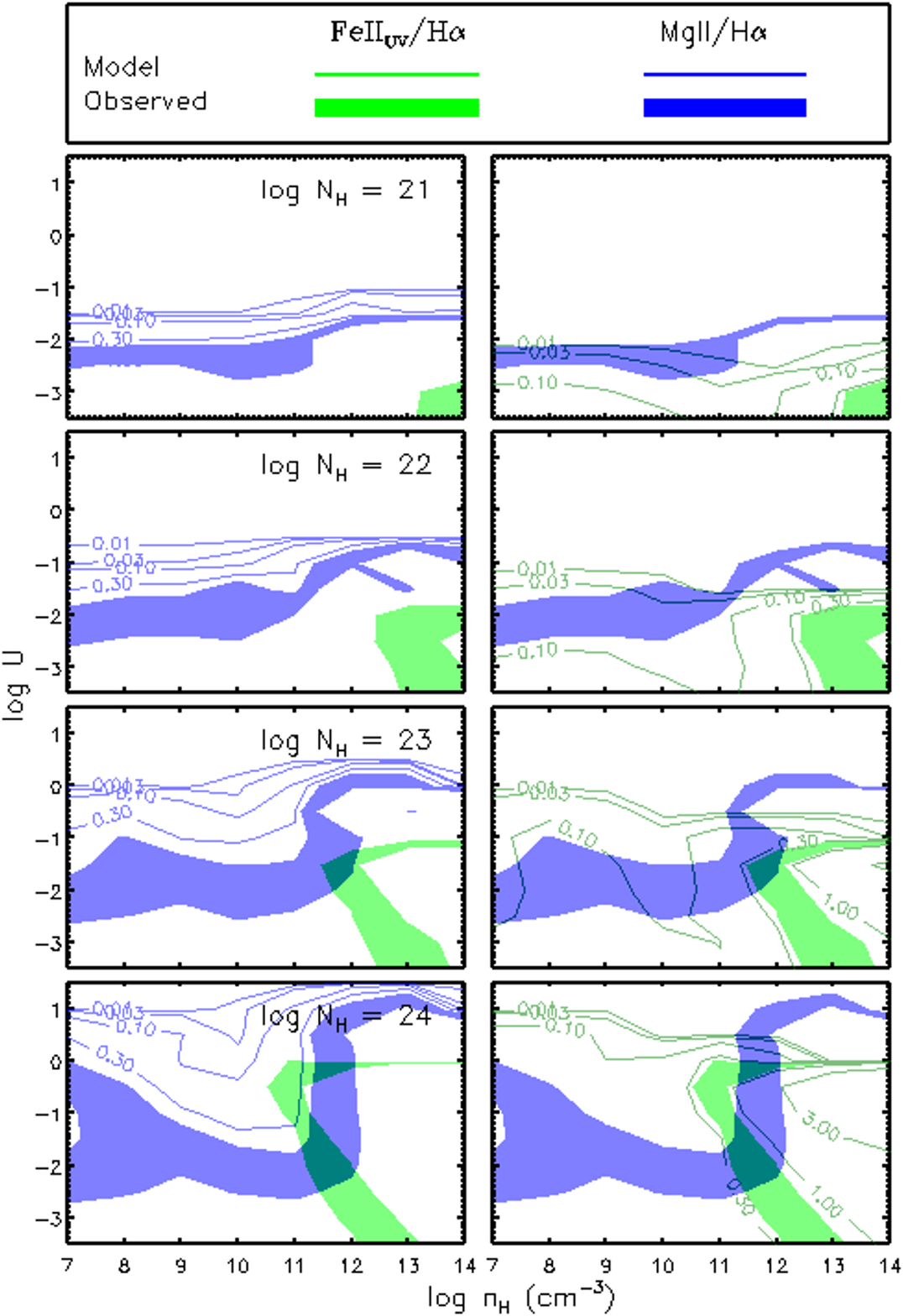}
\caption{Contours of \mgii/\ha\ (blue) and UV \feii/\ha\ (green) as a function
of $n_H$ and U calculated by CLOUDY for the column density $N_H=10^{21}-10^{24}$ cm$^{-2}$, solar abundance, and MF87 SED.
When the $N_H \geq 10^{23}$ cm$^{-2}$, the $\rm 1-\sigma $ confidence levels of \mgii/\ha\ and UV \feii/\ha\ start to overlap.
$N_H = 10^{23}$ cm$^{-2}$ can be considered as the lower limit on column density of the outflow gas.}
\label{f8}
\end{figure*}

\begin{figure*}[ht]
\epsscale{1}
\plotone{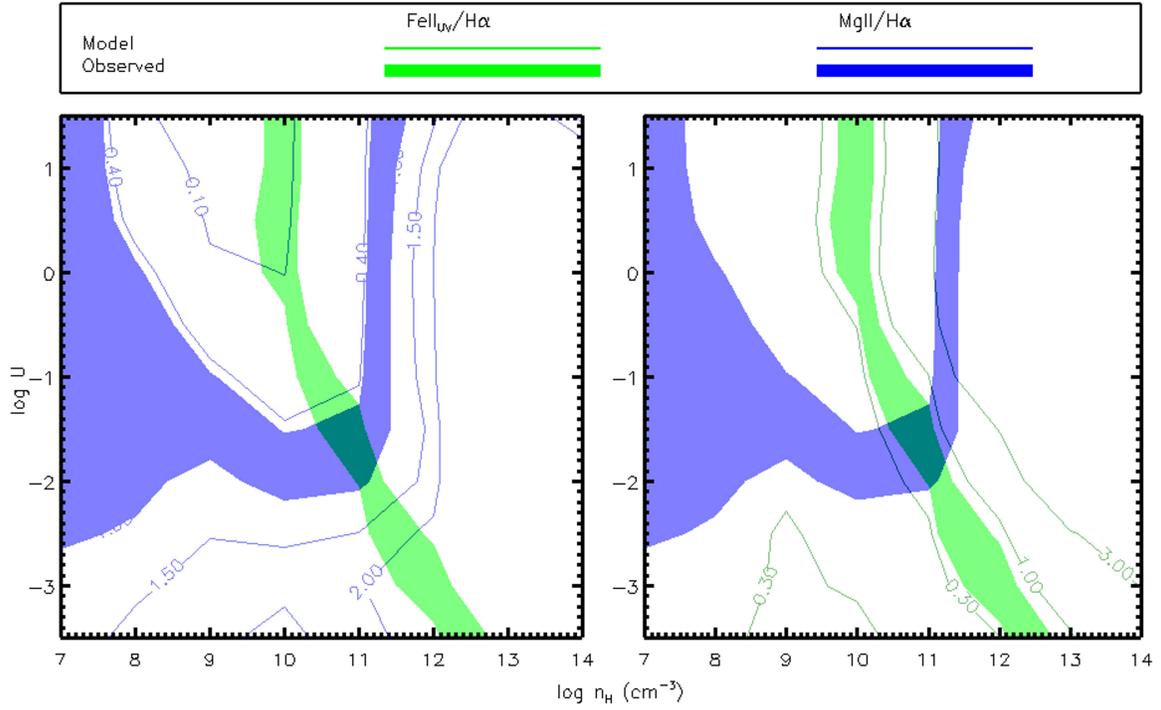}
\caption{Contours of \mgii/\ha\ (blue) and UV \feii/\ha\ (green) as a function
of $n_H$ and U. The calculations are same as Figure 8, but assuming ionization boundary.
The overlapped region constrains the parameters of outflow gas to a narrow region of $n_H$ from $10^{10.6}$ to $10^{11.3}$ cm$^{-3}$ and log U from -2.1 to -1.5.}
\label{f9}
\end{figure*}

\begin{figure*}[ht]
\epsscale{0.8}
\plotone{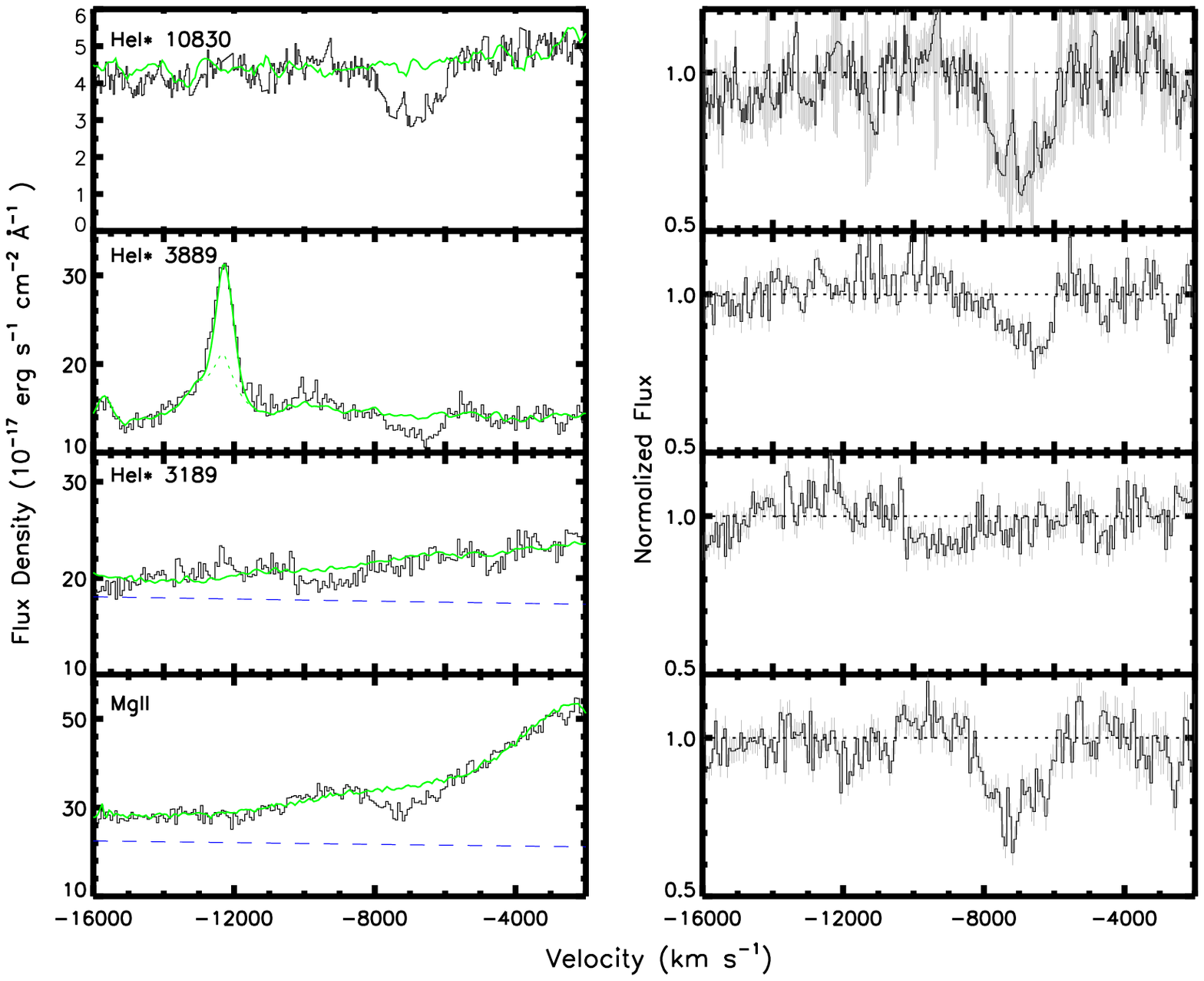}
\caption{The pair matching results and normalized spectra of \mgii, \heitoen, \heiteen\ and \heiozetz.
The pair matching results can be directly considered as the absorption-free spectra of \heiteen. For \heiozetz, based on the conclusion that the absorption component partially obscures the accretion disk, the radiation from the hot dust near \heiozetz\  has been subtracted.
For the same reason, the continuum from the accretion disk is chosen as the absorption free spectra of \mgii\ and \heitoen.}
\label{f10}
\end{figure*}

\begin{figure*}[ht]
\epsscale{0.6}
\plotone{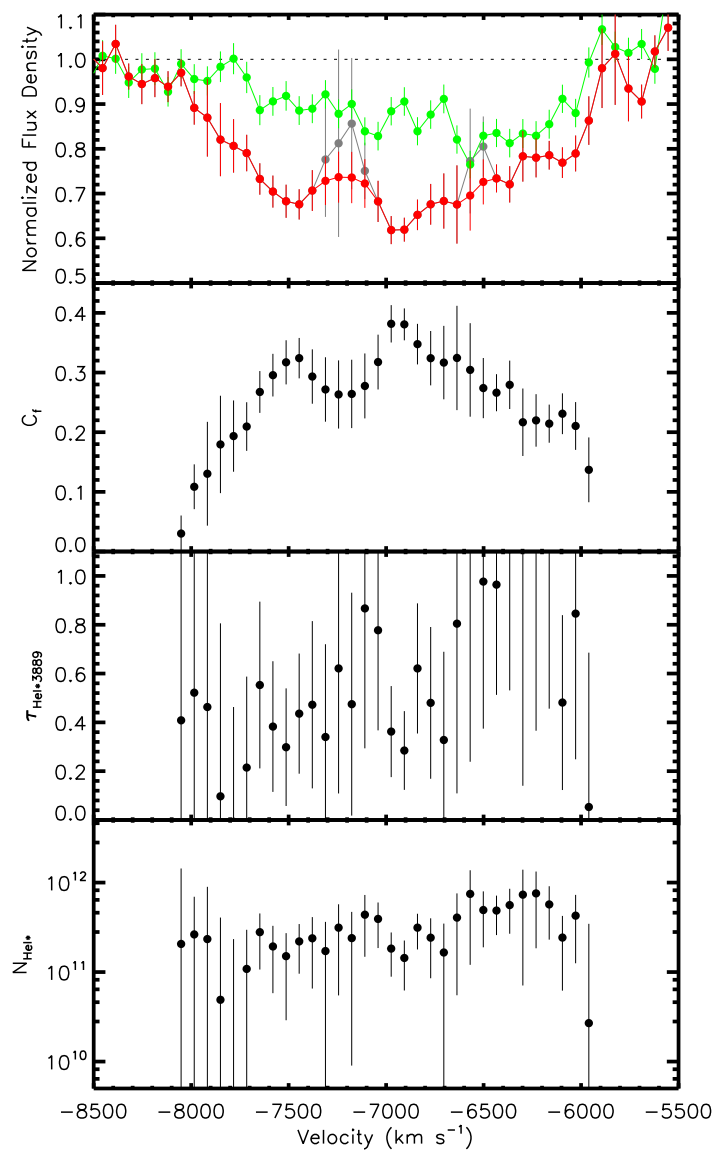}
\caption{ Normalized absorption spectrum of \heiteen\ (green) and \heiozetz\ (red).
After masking the pixels seriously affected by sky lines (gray), the equation 1 is employed at each pixel to
obtain the $C_f$ and $N_{col}$ of \hei. The typical $C_{f}$ is around 0.3 and the integral $N_{col}$ of \hei\ is $(5.0 \pm 0.7) \times 10^{14}$ cm$^{-2}$.}
\label{f11}
\end{figure*}

\begin{figure}[ht]
\epsscale{0.6}
\plotone{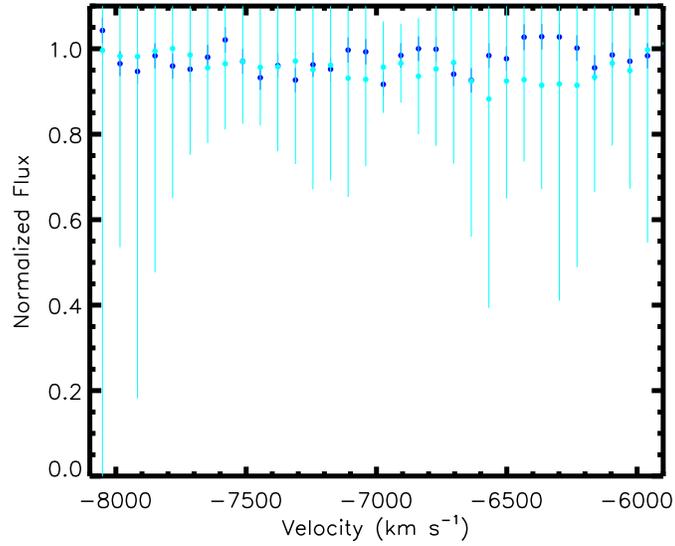}
\caption{The comparison between the simulated \heitoen\ trough (cyan) and observed \heitoen\ trough (blue). The consistence of the two
indicates that the derived $C_f$ and $N_{col}$ of \hei\ are reliable  }
\label{f12}
\end{figure}

\begin{figure}[ht]
\epsscale{0.6}
\plotone{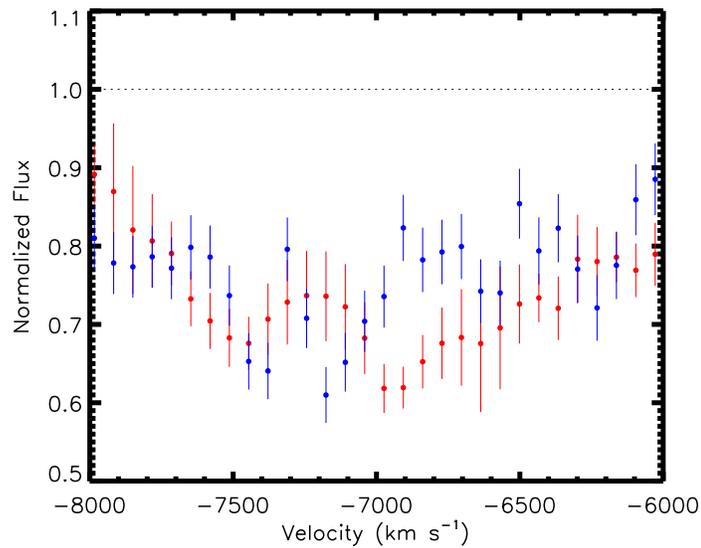}
\caption{The comparison between the \mgii\ BAL trough (red) and 1-$C_f$ (blue). It indicates the saturation of \mgii\ absorption trough.  }
\label{f13}
\end{figure}

\begin{figure}[ht]
\epsscale{0.8}
\plotone{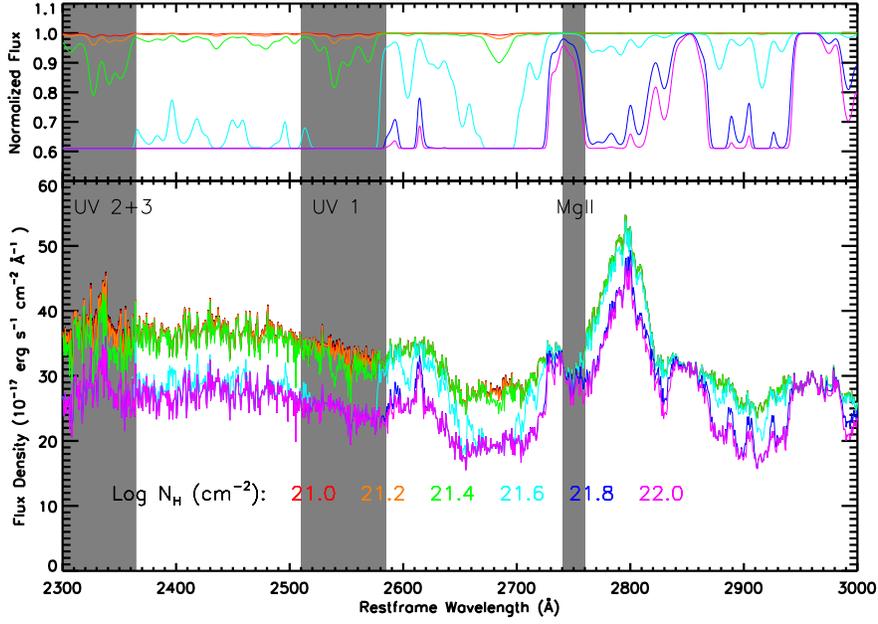}
\caption{{\it Upper panel:} The modelled absorption profile of UV \feii\ in the wavelength range of 2300 to 3000 \AA.
The density of BAL outflow in the model is set to be log n$\rm _{H}$ = 11 and the ionization parameter is log U = -1.9.
A series of column density (N$\rm _H$) is employed in the model from 10$^{21}$ to 10$^{22}$ with a dex step of 0.2.
For clarity, we only show the model with logN$\rm _H$=21.0, 21.2, 21.4, 21.6, 21.8 and 22.0, as color-coded curve.
The absorption profile of single \feii\ line is assumed to be the same with that of \hei.
{\it Lower panel:} The observed spectrum is shown in black. We also added each modelled spectrum above to the observed spectrum
for comparison.
}
\label{f14}
\end{figure}

\begin{figure}[ht]
\epsscale{1}
\plotone{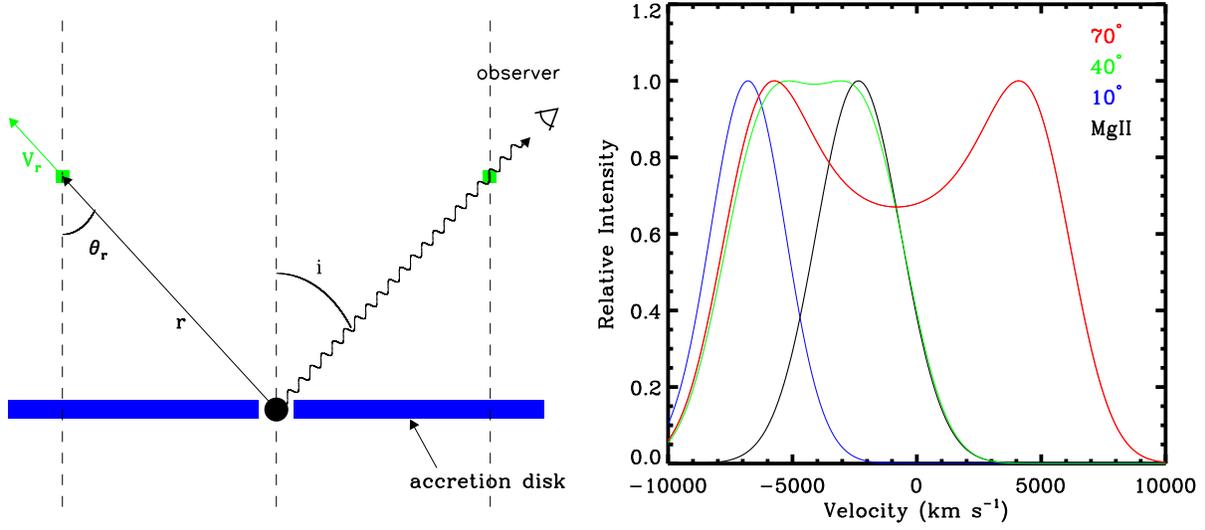}
\caption{\textbf{Left}: The cross-section of the ring model. The angle relative to the normal line of the accretion disk is $ \theta_r $. The distance of the outflow is estimated at r = 9000$r_g$. The radial velocity $v_r$ = 7000\kmps. The angle for the line of sight is $i$(=$\theta_r$). Only one parameter $\theta_r$ is free in this model. \textbf{Right}: Comparisons of the three modelled results ($\theta_r$ = 10$^\circ$, 40$^\circ$, 70$^\circ$) with the fitted blueshifted \mgii\ profile. }
\label{f15}
\end{figure}

\begin{figure}[ht]
\epsscale{1}
\plotone{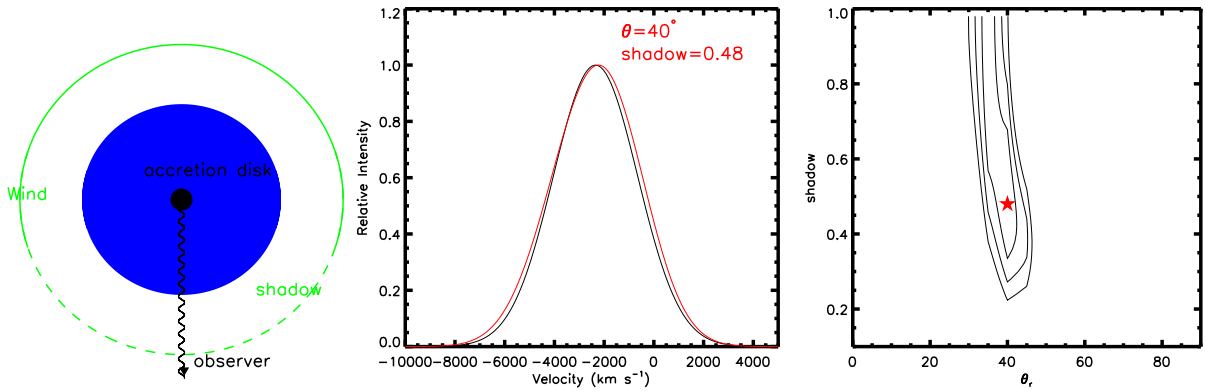}
\caption{\textbf{Left}: The top view  of the proposed "shadow" model. The setting is the same with the ring model except for
the addition of a free parameter "shadow". \textbf{Middle}: Comparisons of the best-fitted result ($\theta_r$ = 40$^\circ$, shadow = 0.48)
with the blueshifted \mgii. \textbf{Right}: The 1, 2, and 3$\sigma$ confidence levels for $\theta_r$ versus the shadow parameter.
Red star denotes the best-fitted value.
}
\label{f16}
\end{figure}

\clearpage

\appendix

\section{Multi-column density modelling of the outflow emission line}

It should be noted that CLOUDY photoionization simulations in this paper (also many other works in literature)
are based on the assumption that the density and column density are uniformly distributed in the specific outflows. However, this model may be over-simplified.
According to the outflow models in Proga et al. (2000) and Higginbottom et al. (2014), the density and column density
of the outflow can be variable with different locations and directions. However, an outflow model with multiple densities and
column densities may be too complex to be constrained by the observations of SDSS J1633+5127.
If assuming that the BEL and BAL outflows have the similar density, the outflows in SDSS J1633+5127 can be
simplified to a slab-shaped medium with multiple column densities and uniform density. In this model,
the low-ionization blueshifted BELs, such as \mgii, \feii, tend to trace the outflow gas with higher column
density. In addition, if further assuming that the covering factor of the outflow gas is related to the column density,
i.e., covering factor decreases as the column density increases, the line of sight would have greater chance to peer
through the outflow gas with lower column density. This may explain why the column density derived
from the blueshifted BELs is higher than that from BALs.
In order to further constrain the outflow properties of SDSS J1633+5127, we attempted to reproduce
the \mgii\ BEL profile with this outflow model.   For this model, the \mgii\ BEL profile can be expressed as 

   \begin{equation}
     \Psi=\int F(N_{H})\psi(N_{H})C_{f}^{'}(N_{H}) d(N_{H})
   \label{functions:profile}
   \end{equation}
, where the F(N$\rm _{H} $) is the intensity of the outflow gas at a given N$\rm _{H} $. The $ \psi$(N$\rm _{H}$) is the profile of the outflow gas with specific N$\rm _{H} $ caused by the geometry of the outflow. The C$ _{f}^{'}$(N$\rm _{H}$) dN$\rm _{H}$ is the covering factor of the outflow gas in the range of N$\rm _{H} $ to N$\rm _{H}$+dN$\rm _{H} $, which is assumed to be proportional to N$\rm _{H}^{-\Gamma}$.

\begin{figure}[ht]
\epsscale{1}
\plotone{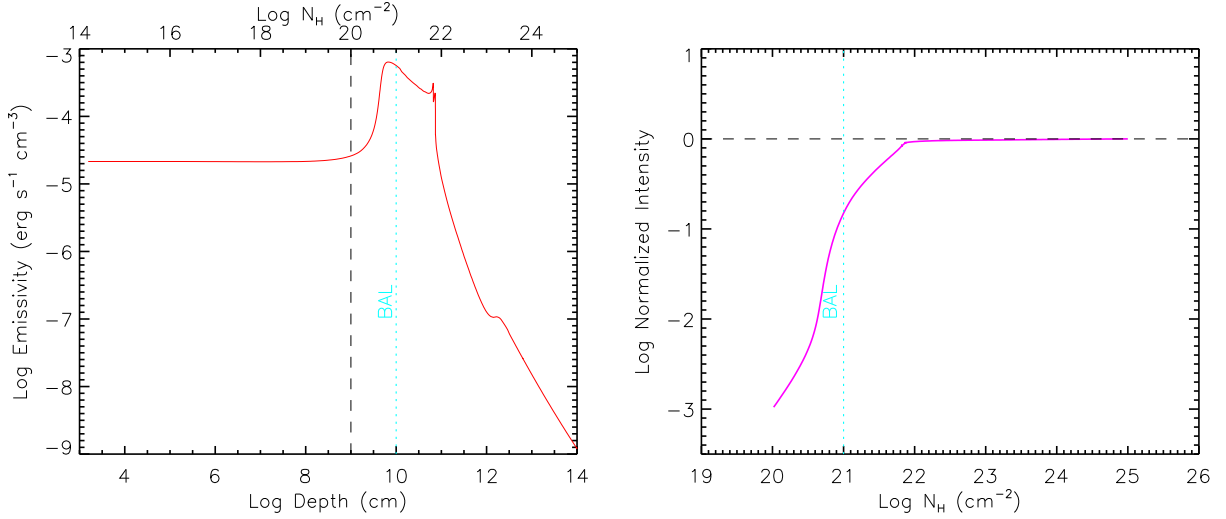}
\caption{\textbf{Left}: The \mgii\ emergent emissivity varies with the ionized depth in the case of the density n$ _{H} $ = 10$ ^{11} $ cm$ ^{-3} $ and the ionization parameter log U =-1.9. The corresponding column density of the ionized depth is also shown. We select the column density range of 10$ ^{20} $ to 10$^{25}$ cm$ ^{-2} $ to contain the main emitting region of \mgii.   \textbf{Right}: The \mgii\ intensity vs. the column density in the range of 10$ ^{20} $ to 10$^{25}$ cm$ ^{-2} $ in the case of  n$ _{H} $ = 10$ ^{11} $ cm$ ^{-3} $ and log U =-1.9. The intensity is normalized by the maximum.    }
\label{f17}
\end{figure}

Using the density and ionization parameter derived for the BEL outflow, log n$\rm _{H} $ (cm$ ^{-3} $) = 11 and log U = -1.9,
we can obtain the \mgii\ emergent emissivity distribution along with the ionized depth via CLOUDY simulations.
The result is displayed in the left panel of Fig.\ref{f17}.
The distribution indicates that the \mgii\ emission can be ignored when the column density is lower than 10$ ^{20} $ cm$ ^{-2} $. Therefore,
we set 10$ ^{20} $ cm$ ^{-2} $ as the lower limit for column density of the multi-column density outflow model.
The upper limit on the column density is set to be 10$ ^{25} $ cm$ ^{-2} $.
We then derived the \mgii\ intensity (F(N$\rm _{H} $)) as a function of column density,
which is shown in the right panel of Fig.\ref{f17}. 

\begin{figure}[ht]
\epsscale{1}
\plotone{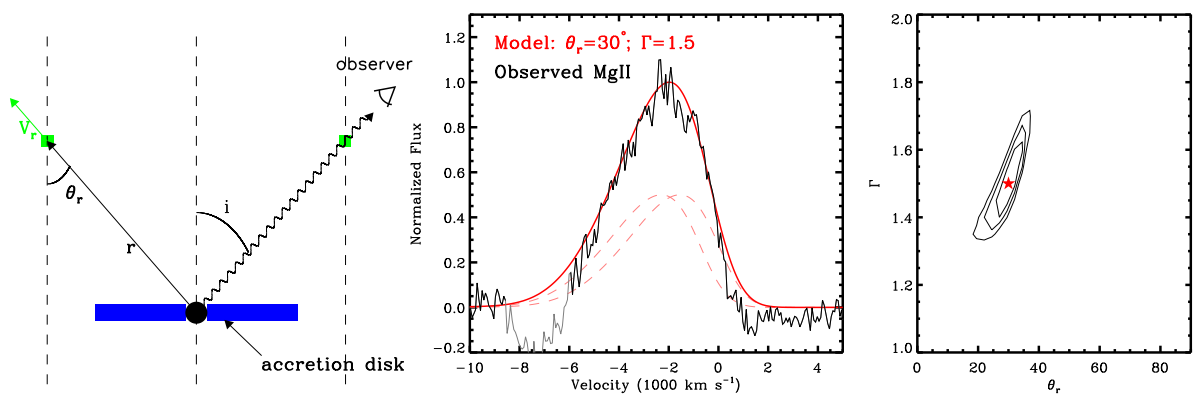}
\caption{\textbf{Left}: The cross-section of the ring model. The angle relative to the normal line of the accretion disk is $ \theta_r $. The angle for the line of sight is $i$(=$\theta_r$).  \textbf{Middle}: Comparisons of the best-fitted result ($\theta_r$ = 30$^\circ$, $ \Gamma $ = 1.5)
with the blueshifted \mgii\ observed profile. The modelled \mgiis\ and \mgiie\ are plotted in pink dashed lines and the total profile of \mgii\ doublet is displayed in red solid line. The mis-match between the model and observation at about -7000 \kmps is due to the \mgii\ BAL, where the observed \mgii\ profile is marked in gray.   \textbf{Right}: The 1, 2, and 3$\sigma$ confidence levels for $\theta_r$ versus the $ \Gamma $ parameter.
Red star denotes the best-fitted value. }
\label{f18}
\end{figure}

While a biconical structure is always considered as the geometry of outflows in previous works (e.g. Elvis 2000),
it is two-dimensional and need a certain number of free parameters to reproduce the emission line profile. For simplicity,
 we employed a one-dimensional "ring" model to reproduce the blueshifted \mgii\ profile of
 SDSS J1633+5137. The cross-section of this model is displayed in the left panel of Fig.\ref{f18}.

The ring model for the outflow assumes that the line originates on a ring above the disk
 for which axis inclines with an angle $ i $ relative to the line of sight.
 The ring has an angle $ \theta_r $ relative to the normal direction of the accretion disk.
 For SDSS J1633+5137, as the blueshifted BELs and BALs of the outflow are observed at the same time,
 it is natural to assume our line of sight is penetrating through the outflow,
 which means the angel $i$ = $ \theta_r $. The distance from the ring to the black hole is $r$ (expressed in units of the gravitational radius, $r_{g}$). The distance of outflow to the central source derived from blueshifted BELs is about 0.1pc (or 9000 $r_{g}$).  The outflow velocity along with the radial direction is $v_r$.  The coordinate of gas in the ring can be
 expressed as( $r, \theta_r, \phi $), where the $\phi$ changes from -$\pi$ to $\pi$ and our line of sight
 corresponds to $\phi$ = 0. Assuming the outflowing ring's rotation increases with the direction of $\phi$,
 for a certain ring at $r, \theta_r$, and $\phi $, the velocity on the line of sight $v_{obs}$ can be expressed
 as:

    \begin{equation}
     v_{obs}(r,\theta_r, \phi)=-(v_rsin^2(\theta_r)cos(\phi)+v_rcos^2(\theta_r)).
   \label{functions:vobs}
   \end{equation}

 To compare with the observed profile easily, we defined the direction of far away the central BH as
the positive direction of $v_{obs}$.  With the specific $ v_r $ and $ \theta_r $, we can derive the emission profile of the ring.
According to the outflow theory, if an outflow is only regulated by gravitation and ionizing radiation,
the $ v_r $ can be considered proportional to N$\rm _{H}^{-0.5} $ (Netzer \& Marziani 2010; Marziani et al. (2013)).
Assuming the BAL outflow is physically connected with the BEL, we can use the $ v_r $ of BAL outflow gas (-7000 \kmps)
and the column density (10$^{21.2}$ cm$ ^{-2}$) to
constrain the scaling factor for the above relation.
Based on this, we calculated the $ v_r$(N$\rm _{H}$)
for different values of N$\rm _{H} $. With this $ v_r$(N$\rm _{H}$) and Eq.\ref{functions:vobs}, we can derive the emission
profile $\psi$(N$\rm _{H}$). This profile includes a free parameter $ \theta_r $.  In addition to $\psi$(N$\rm _{H}$), the C$ _{f}^{'} $ relative value can be obtained through the relationship of C$ _{f}^{'}$ (N$\rm _H$) $ \propto $  N$\rm _{H}^{-\Gamma} $ by introducing
another free parameter $ \Gamma $.
Thus, together with the F(N$\rm _{H} $) as determined by the photoionization simulations (Fig. 17, right), the emission profile
of the multi-column density outflow model can be derived.   

In the middle panel of Fig.\ref{f18}, we display a modelled profile which can reproduce the
profile of \mgii\ well. The parameters of this best-fitted profile are
$\theta_r$ = 30$^\circ$ and $ \Gamma $ = 1.5.
Note that the free parameter $\theta_r$ and $\Gamma$ can be well constrained
in our model.
Fig.\ref{f18} (right panel) shows the 1, 2, and 3$\sigma$ confidence levels of the parameter $\theta_r$
versus $\Gamma$.
At 1$\sigma$ confidence level, the $\theta_r$ was constrained in the range from 25$^\circ$ to 35$^\circ$, while
the $\Gamma$ was from 1.4 to 1.65.
This multi-column density outflow model may explain the difference in the column density of
the BAL and BEL gas for the source SDSS J1633+5127 if they are physically connected.
The inferred model parameters will be also valuable for the simulations of AGN outflows
with multiple column density distributions.

\end{document}